\documentclass[twocolumn,times]{aastex62}

\usepackage{natbib}
\usepackage{placeins}
\usepackage{amsmath}
\usepackage{mathtools}
\newcommand\degree{{^\circ}}

\newcommand\kms{km~s$^{-1}$}
\defcitealias{walker2015}{W15}
\defcitealias{kleyna2002}{K02}
\defcitealias{kleyna2003}{K03}
\defcitealias{wilkinson2004}{W04}
\defcitealias{kirby2010}{K10}
\defcitealias{olszewski1995}{O95}
\defcitealias{armandroff1995}{A95}
\defcitealias{spencer2017b}{Paper I}
\defcitealias{duquennoy1991}{DM91}
\defcitealias{raghavan2010}{R10}
\defcitealias{fischer1992}{FM92}
\defcitealias{marks2011}{MK11}

\begin{document}

\title{\large\textbf{The Binary Fraction of Stars in Dwarf Galaxies: the Cases of Draco and Ursa Minor}}
\shorttitle{Binary Fraction of Dra and UMi}
\shortauthors{Spencer et al.}
\correspondingauthor{Meghin Spencer}
\email{meghins@umich.edu}

\author{Meghin E. Spencer}
\affil{Department of Astronomy, University of Michigan, Ann Arbor, MI, USA}

\author{Mario Mateo}
\affil{Department of Astronomy, University of Michigan, Ann Arbor, MI, USA}

\author{Edward W. Olszewski}
\affil{Steward Observatory, The University of Arizona, Tucson, AZ, USA}

\author{Matthew G. Walker}
\affil{McWilliams Center for Cosmology, Department of Physics, Carnegie Mellon University, Pittsburgh, PA, USA}

\author{Alan W. McConnachie}
\affil{NRC Herzberg Institute of Astrophysics, Victoria, BC, Canada}

\author{Evan N. Kirby}
\affil{Department of Astronomy, California Institute of Technology, Pasadena, CA, USA}

\begin{abstract}
Measuring the frequency of binary stars in dwarf spheroidal galaxies (dSphs) requires data taken over long time intervals. We combine radial velocity measurements from five literature sources taken over the course of $\sim30$ years to yield the largest multi-epoch kinematic sample for stars in the dSphs Draco and Ursa Minor. With this data set, we are able to implement an improved version of the Bayesian technique described in \citet{spencer2017b} to evaluate the binary fraction of red giant stars in these dwarf galaxies. Assuming \citet{duquennoy1991} period and mass ratio distributions, the binary fractions in Draco and Ursa Minor are $0.50^{+0.04}_{-0.06}$ and $0.78^{+0.09}_{-0.08}$, respectively. We find that a normal mass ratio distribution is preferred over a flat distribution, and that log-normal period distributions centered on long periods ($\mu_{\log P}>3.5$) are preferred over distributions centered on short ones. We reanalyzed the binary fractions in Leo II, Carina, Fornax, Sculptor, and Sextans, and find that there is $<1\%$ chance that binary fraction is a constant quantity across all seven dwarfs, unless the period distribution varies greatly. This indicates that the binary populations in Milky Way dSphs are not identical in regard to their binary fractions, period distributions, or both. We consider many different properties of the dwarfs (e.g. mass, radius, luminosity, etc.) and find that binary fraction might be larger in dwarfs that formed their stars quickly and/or have high velocity dispersions.
\end{abstract}


\section{Introduction}

Within the solar neighborhood, there are approximately one to two times as many binary star systems as single stars \citep{duquennoy1991,raghavan2010}. The presence of binary systems is also expected within dwarf spheroidal galaxies (dSphs), but the quantity is largely unknown. If the fraction is similar to the solar neighborhood, then the additional radial velocity components of the binary systems can inflate the observed velocity dispersion in some dSphs, which can impact inferences that draw upon the kinematics, such as mass estimates. This effect can be corrected if the attributes of the binary population---including binary fraction and orbital parameter distributions like period, mass ratio, and eccentricity---are well measured. Measurements of the binary populations are also helpful in predicting the frequency of type Ia supernova \citep[e.g.,][]{maoz2017} and in putting constraints on star formation processes in dSphs \citep[][and references therein]{duchene2013}.

A recent, detailed binary analysis has been performed on Leo II \citep{spencer2017b}, Carina, Fornax, Sculptor, and Sextans \citep{minor2013}. Two of the remaining classical Milky dSphs, Draco and Ursa Minor, were well studied in the early ages of individual dSph stellar kinematics \citep{armandroff1995,olszewski1996,hargreaves1996a}, but their binary populations have not been reviewed for the last two decades. During that time, a number of new radial velocities have been obtained for large samples of stars in Draco and Ursa Minor \citep{kleyna2002,kleyna2003,wilkinson2004,kirby2010}. This offers an opportunity to revisit our knowledge of the binary populations in these galaxies. While it is unlikely that binaries will significantly alter our view of the dark matter content in classical dSphs, such as Draco and Ursa Minor, the issue remains open-ended for the more recently discovered ultra-faints. 

Ultra-faints have exhibited much smaller dispersions \citep{simon2007}, and while they are still believed to be dark matter dominated, the role of binaries might be significant in these cases. Velocity contributions from binaries on the order of a few \kms{} are similar to the velocity dispersions of ultra-faints and can act to widen the observed dispersions. It has been shown that binaries are unlikely to inflate the observed dispersions of dwarfs with $\sigma_{obs}>4$ \kms{} (i.e. classicals) by more than 30\% \citep{minor2010}, but that binaries have up to a 40\% chance of boosting dispersions from near-zero values to what is presently observed in dwarfs with $\sigma_{obs}\lesssim4.5$ \kms{} \citep[i.e., ultra-faints; ][]{mcconnachie2010}. Other simulations have shown that dwarfs with intrinsic dispersions of 1 \kms{} can be inflated by a factor of four for populations extremely rich in binaries \citep{spencer2017b}. The severity of the effect varies among simulations due to the details of the velocity samples and the shape of the distributions for the binary orbital parameters. However, every case agrees that dwarfs with low velocity dispersions have a high risk of being inflated by binaries.

Large, multi-epoch kinematic surveys are needed to correct observed velocity dispersion for the inflation caused by binaries on a case by case basis, as was done for Segue 1 \citep{martinez2011,simon2011} and Bootes I \citep{koposov2011}. Unfortunately, such data are not currently available for most ultra-faints. An alternative that can at least explore the range in severity of binaries on the velocity dispersion is to model the effects for different assumptions of the binary fraction and binary orbital parameters. The test parameters can be narrowed down by considering the values occupied by classical dSphs. Properties of the binary populations in classical dSphs are interesting in their own right, but they are also useful in determining how adversely binaries are impacting the velocity dispersions in ultra-faints. 

In this paper, we aim to better constrain the binary fractions in both Draco and Ursa Minor. We will then use this result plus those from other classical dwarfs to estimate the binary fraction in ultra-faints and provide examples on the severity of the effect for ultra-faints. We explore whether or not binary fraction is a constant quantity across all dSphs and also comment on which of our tested period and mass ratio distributions provide the best fit to the data. Section \ref{sec_data} describes the data that we used---including the presentation of a new spectroscopic data set for Ursa Minor---and Section \ref{sec_methods} details our methodology for finding the binary fraction. Our results are in Section \ref{sec_results} and the summary and conclusions are in Section \ref{sec_conclusions}.

\section{Velocity Data}\label{sec_data}

Our analysis aims to define the binary fractions in Draco and Ursa Minor via the presence of velocity variability among the stars. Data must meet several criteria to be used in this analysis. 
\begin{itemize}
\item The stars must be red giants. We make simplifications later about the mass and period distributions for binary stars by assuming that the primaries are red giants. The same assumptions would not be true for main sequence or horizontal branch stars.
\item The stars must be members of the dSphs. Binaries are found in both dSphs and the MW halo, but the frequency with which they are found is not necessarily the same. Since we aim to find the binary fraction specifically within dSphs, we do not want MW halo stars to skew the results.
\item The available velocities cannot be averaged over multiple observing epochs. Doing so would conceal the signatures of velocity variability, which are key in our method of determining the binary fraction.
\item The velocity errors must reflect the measurement uncertainty. As we will see in Section \ref{sec_velerrors}, poorly determined errors can increase or decrease the significance of velocity variation, and thereby lead to incorrect measurements of the binary fraction.
\item The stars must have multi-epoch observations. Velocity variability is identified as a function of time, so we require multiple observations. 
\end{itemize}

There are six data sets each for Draco and Ursa Minor that meet our criteria. These are summarized in Table \ref{ch4_table_datasets}. Column 1 lists the paper order as it appears in Section \ref{sec_datasets}, column 2 lists the reference paper, column 3 lists the data set abbreviation, column 4 lists the number of stars in the data set that adhere to the first four of the above criteria, column 5 lists the median velocity error of those stars, column 6 lists the years when the observations were taken, column 7 lists the number of stars from the data set that we use in this analysis, and column 8 lists the velocity offset that we apply to put the stars on the same velocity standard. In this section we will first introduce the data sets (Section \ref{sec_datasets}) and then describe the ways in which we trimmed them to meet our requirements (Sections \ref{sec_membership}--\ref{sec_velerrors}).

\subsection{Data Sets}\label{sec_datasets}

\begin{deluxetable*}{c l c c c c c c}[t]
\centering 
\tabletypesize{\scriptsize}
\tablewidth{0pt}
\tablecaption{Papers with radial velocity data in Draco and Ursa Minor\label{ch4_table_datasets}}
\tablehead{\colhead{Paper} & \colhead{Paper}  & \colhead{Abbreviation} & \colhead{N Stars} & \colhead{Median $\sigma_v$} & \colhead{Years} & \colhead{N Stars} & \colhead{$v_{\mathrm{offset}}$} \\
\colhead{Number} & & & \colhead{Criteria} & \colhead{(\kms{})} & & \colhead{Usable} & \colhead{(\kms{})}}
\startdata
Draco\\
\hline 
1 & \citet{olszewski1995} & O95 & 20 & 1.8 & 1982-1991 & 20 & -0.41\\
2 & \citet{armandroff1995} & A95 & 86 &  4.2 & 1992-1994 & 75 & -0.41\\
3 & \citet{kleyna2002} & K02 & 158 & 1.7 & 2000 & 140 & -0.17\\
4 & \citet{wilkinson2004} & W04 & 114 & 2.5 & 2003 & 96 & -0.17\\
5 & \citet{kirby2010} & K10 & 305 & 2.5 & 2009 & 123 & 0.21\\
6 & \citet{walker2015} & W15 & 414 & 0.9 & 2006-2011 & 292 & 0.0\\
\hline 
Ursa Minor \\
\hline 
1 & \citet{olszewski1995} & O95 & 16 & 1.9 & 1983-1989 & 16 & 0.06\\
2 & \citet{armandroff1995} & A95 & 90 & 4.3 & 1992-1994 & 88 & 0.06\\
3 & \citet{kleyna2003} & K03 & 64 & 5.1 & 2002 & 58 & -1.07\\
4 & \citet{wilkinson2004} & W04 & 146 & 2.9 & 2003 & 112 & -1.07\\
5 & \citet{kirby2010} & K10 & 336 & 2.4 & 2009-2010 & 136 & -0.24\\
6 & (Table \ref{tab:umi_table1}) & Tab\ref{tab:umi_table1} & 404 & 1.0 & 2008-2011 & 250 & 0.0\\
\enddata 
\end{deluxetable*}

The first data set is \citet[][hereafter O95]{olszewski1995}. Using the echelle spectrograph on the Multiple Mirror Telescope, they measured velocities every year between 1982 and 1991. They collected data for 24 stars in Draco and 18 stars in Ursa Minor. Subsets of this data were presented in \citet{aaronson1987,aaronson1988} and \citet{olszewski1988}.

The second data set \citep[][hereafter A95]{armandroff1995} was obtained with the Hydra multi-fiber positioner and the Bench Spectrograph on the KPNO 4-meter telescope. They observed many of the same stars as \citetalias{olszewski1995} in both Draco and Ursa Minor during the years 1992--1994. The sample expanded greatly to include 91 stars in Draco and 94 in Ursa Minor. 

The third data set was split into two papers, with \citet[][hereafter K02]{kleyna2002} focusing on Draco and \citet[][hereafter K03]{kleyna2003} focusing on Ursa Minor. They used the AF2/WYFFOS fiber-fed spectrograph on the William Herschel Telescope during the year 2000 for Draco and 2002 for Ursa Minor. 

The fourth data set \citep[][hereafter W04]{wilkinson2004} is a follow-up to the previous \citetalias{kleyna2002} and \citetalias{kleyna2003} data, using the same instrument and telescope. They took measurements in the year 2003 to compose a second epoch of data for about a third of the Kleyna stars in Draco and about one half of the Kleyna stars in Ursa Minor. 

The fifth set of data was described in \citet[][hereafter K10]{kirby2010}. They observed stars in both dwarfs during 2009 using Keck/DEIMOS. Additional stars in Ursa Minor were observed in 2010, although these were not published. Each star only received a single epoch of observations, but many of the stars appeared in other data sets, making them useful to our research.

The sixth data set comes from MMT/Hectochelle observations of Draco and Ursa Minor during the years 2006 - 2011.  The Draco data are already published \citep[][hereafter W15]{walker2015}, and we present the Ursa Minor data in Section \ref{subsubsec:umi_data}.  

There were three other studies with radial velocities of red giants in Draco and/or Ursa Minor, but these failed to meet one or more of the criteria listed in Section \ref{sec_data}. \citet{jardel2013} observed 13 stars in Draco during a single epoch, but only one star exists in the other data sets. This sample does not appreciably add to the size of the combined data or expand the temporal information. It would also be impossible to put it on the same velocity standard as the other data, given the minimal overlap. (This step is described in Section \ref{sec_velstandard}.)

\citet{munoz2005} used Keck HIRES to obtain radial velocities of 52 stars in Ursa Minor over two epochs separated by 2 years. These were later supplemented with additional observations of both Ursa Minor and Draco. Only the average velocities from this observing program were available, so these data could not be used in our analysis.

Lastly, \citet{hargreaves1994b,hargreaves1996b} published velocity data for Ursa Minor and Draco, respectively. It was found that the velocity errors of these data were systematically underestimated by about 15\%, an effect likely caused by poor sky subtraction \citep{armandroff1995}. Underestimated velocity errors would artificially increase the binary fraction that we measure; therefore we have chosen to exclude this data set from our analysis.

\subsubsection{New MMT/Hectochelle Observations of Stars in Ursa Minor}\label{subsubsec:umi_data}

\begin{table*}[t]
\centering
\scriptsize 
\caption{new MMT/Hectochelle data from individual observations of RGB candidates in Ursa Minor$^{a}$}
\begin{tabular}{@{}lrccccccccccccccccccccc@{}}
\hline 
$\alpha_{2000}$&$\delta_{2000}$&HJD&S/N$^{b}$&$\overline{v_{\rm los}}$&$\overline{T_{\rm eff}}$&$\overline{\log_{10}g}$&$\overline{\mathrm{[Fe/H]}}$\\
(hh:mm:ss)&($^{\circ}$:$'$:$''$)&(days)&&(km s$^{-1}$)$^{c}$&(K)&(dex)$^{d}$&(dex)\\
\hline 
15:10:55.26&+66:46:53.2&$2454614.87$&$  8.9$&$-175.4\pm 0.5^{( 0.0,3.0)}$&$4991\pm  96^{( 0.1, 3.0)}$&$ 2.0\pm 0.2^{(-0.1, 3.1)}$&$-1.55\pm0.12^{( 0.0, 2.9)}$\\
  & &$2454615.75$&$ 14.0$&$-176.5\pm 0.4^{( 0.1,3.2)}$&$4960\pm  71^{( 0.0, 3.1)}$&$ 2.0\pm 0.2^{(-0.0, 3.0)}$&$-1.41\pm0.09^{( 0.0, 3.1)}$\\
  & &$2454615.83$&$ 13.8$&$-176.3\pm 0.4^{( 0.1,3.1)}$&$4930\pm  66^{( 0.0, 3.1)}$&$ 2.0\pm 0.1^{(-0.1, 3.1)}$&$-1.43\pm0.08^{(-0.0, 3.0)}$\\
  & &$2455232.94$&$  7.1$&$-177.4\pm 0.5^{( 0.0,3.0)}$&$4855\pm  90^{( 0.1, 3.1)}$&$ 1.3\pm 0.2^{(-0.3, 3.1)}$&$-1.37\pm0.11^{( 0.0, 3.0)}$\\ 
15:10:35.69&+66:45:56.2&$2454614.87$&$ 11.8$&$ -57.3\pm 0.5^{( 0.0,2.9)}$&$4828\pm  56^{( 0.1, 3.2)}$&$ 5.1\pm 0.2^{( 0.2, 3.2)}$&$-0.15\pm0.07^{(-0.0, 2.9)}$\\
  & &$2454615.83$&$ 17.6$&$ -57.3\pm 0.4^{( 0.1,3.1)}$&$4829\pm  53^{( 0.1, 3.0)}$&$ 5.3\pm 0.1^{( 0.1, 2.7)}$&$ 0.01\pm0.06^{( 0.0, 3.0)}$\\
  & &$2455232.94$&$  9.4$&$ -57.9\pm 0.4^{( 0.2,3.3)}$&$4789\pm  58^{(-0.1, 2.9)}$&$ 5.2\pm 0.2^{( 0.2, 3.1)}$&$ 0.20\pm0.07^{( 0.1, 3.0)}$\\ 
15:11:01.43&+66:43:19.7&$2455232.94$&$  1.3$&$-302.4\pm 1.8^{(-0.2,3.1)}$&$4897\pm 499^{( 1.6, 6.6)}$&$ 1.7\pm 0.8^{( 1.1, 4.1)}$&$-1.00\pm0.69^{( 0.9, 3.3)}$\\
15:08:29.92&+66:52:20.2&$2454614.87$&$  4.9$&$-248.9\pm 1.5^{( 0.4,3.7)}$&$5110\pm 473^{( 0.7, 3.6)}$&$ 3.6\pm 0.7^{( 0.1, 3.2)}$&$-2.65\pm0.50^{( 0.5, 2.9)}$\\
 & &$2454615.83$&$  8.6$&$-247.1\pm 1.0^{( 0.2,3.0)}$&$4653\pm 193^{( 0.4, 3.0)}$&$ 1.5\pm 0.5^{( 0.5, 2.7)}$&$-2.99\pm0.22^{( 0.5, 3.2)}$\\
 & &$2454915.94$&$  6.2$&$-247.0\pm 1.4^{(-0.0,3.3)}$&$5170\pm 349^{( 0.4, 3.6)}$&$ 1.4\pm 0.6^{( 1.2, 4.8)}$&$-2.54\pm0.34^{( 0.0, 3.3)}$\\
 & &$2455232.94$&$  5.8$&$-245.5\pm 1.2^{( 0.0,3.3)}$&$4536\pm 176^{( 0.8, 3.6)}$&$ 1.2\pm 0.4^{( 0.6, 3.0)}$&$-3.06\pm0.19^{( 0.8, 4.0)}$\\
 & &$2455659.75$&$  7.6$&$-245.7\pm 1.2^{(-0.2,3.2)}$&$5121\pm 478^{( 0.6, 3.3)}$&$ 2.3\pm 0.8^{( 0.1, 2.5)}$&$-2.76\pm0.52^{( 0.4, 2.6)}$\\
\hline 
\end{tabular}
\\
\raggedright
$^{a}$See the electronic edition for a complete data table.\\
$^{b}$median signal-to-noise ratio per pixel\\
$^{c}$line-of-sight velocity in the heliocentric rest frame\\
$^{d}$units of $g$ are cm s$^{-2}$\\
\label{tab:umi_table1}
\end{table*}

Before compiling a broader sample using previous studies, we first we present new data from our spectroscopic observations of individual red giant branch (RGB) candidates in Ursa Minor, acquired using the Hectochelle multi-fiber spectrograph \citep{szentgyorgyi06} at the MMT on Mount Hopkins, Arizona. The observational setup (wavelength range 5160 - 5280\AA, resolving power $\mathcal{R}\sim 20,000$ ) and data reduction procedures are identical to those that \citetalias{walker2015} describe in detail for our previous study of Draco. Using the same spectroscopic modeling procedure described by \citetalias{walker2015}, we fit to each spectrum a model that has free physical parameters specifying line-of-sight velocity $v_{\rm los}$, effective temperature $T_{\rm eff}$, surface gravity $\log g$, and metallicity [Fe/H].  Table \ref{tab:umi_table1} lists these measured quantities, along with equatorial coordinates, heliocentric Julian date of observation, and median signal-to-noise ratio per pixel. Following \citetalias{walker2015}, we list parenthetically for each measured parameter the skewness, $S$, and kurtosis, $K$, of the posterior probability distribution function. These quantities provide objective criteria for quality control---we discard observations for which the posterior PDF for velocity is significantly non-Gaussian, with $|S|>1$ and/or $|K-3|>1$.  

After applying this quality-control cut, the new observations contribute 1407 observations of 973 unique RGB candidates in Ursa Minor, including up to five distinct epochs of observation for individual stars. The minimum, median, and maximum 1$\sigma$ errors for individual velocity measurements are $0.4$, $0.7$, and $3.7$ km s$^{-1}$, respectively. All wavelength-calibrated spectra, as well as our model fits, are available at the Zenodo online database\footnote{http://dx.doi.org/10.5281/zenodo.1413660}.

\subsection{Sample Definition}\label{sec_membership}

Each data set was trimmed to match the needs of this analysis, as described in Section \ref{sec_data}. For \citetalias{olszewski1995} and \citetalias{armandroff1995} we removed four or five carbon stars, as we are only concerned with RGB stars here.

The \citetalias{kleyna2002}, \citetalias{kleyna2003}, and \citetalias{wilkinson2004} data sets contained stars both in the dSphs and in the MW foreground. We removed the halo stars on the basis of radial velocity. Normally, a wide cut in velocity risks including MW members, while a narrow cut risks excluding dSph binaries. However, the location of the cut turned out not to matter much in this case, because most of the stars just inside or just outside this limit were discarded later because they only had one epoch of observations, even after combining with the other data sets. The membership criteria that we used were $-330<v_{mem}<-250$ \kms{} for Draco and $-300<v_{mem}<-200$ \kms{} for Ursa Minor. There is still some possible contamination from the Milky Way, but we expect this to only be a couple stars. The effect of such contamination on our results will be negligible.

For \citetalias{kirby2010}, measurements with velocity errors larger than 10 \kms{} were discarded. Velocity nonmembers were present in the Draco data, so we removed likely foreground stars that had velocities less than -320 \kms{} or greater than -265 \kms{}. These limits are the same as what we applied to the \citetalias{walker2015} data set, which is described later.

\begin{figure*}
\epsscale{1}
\plottwo{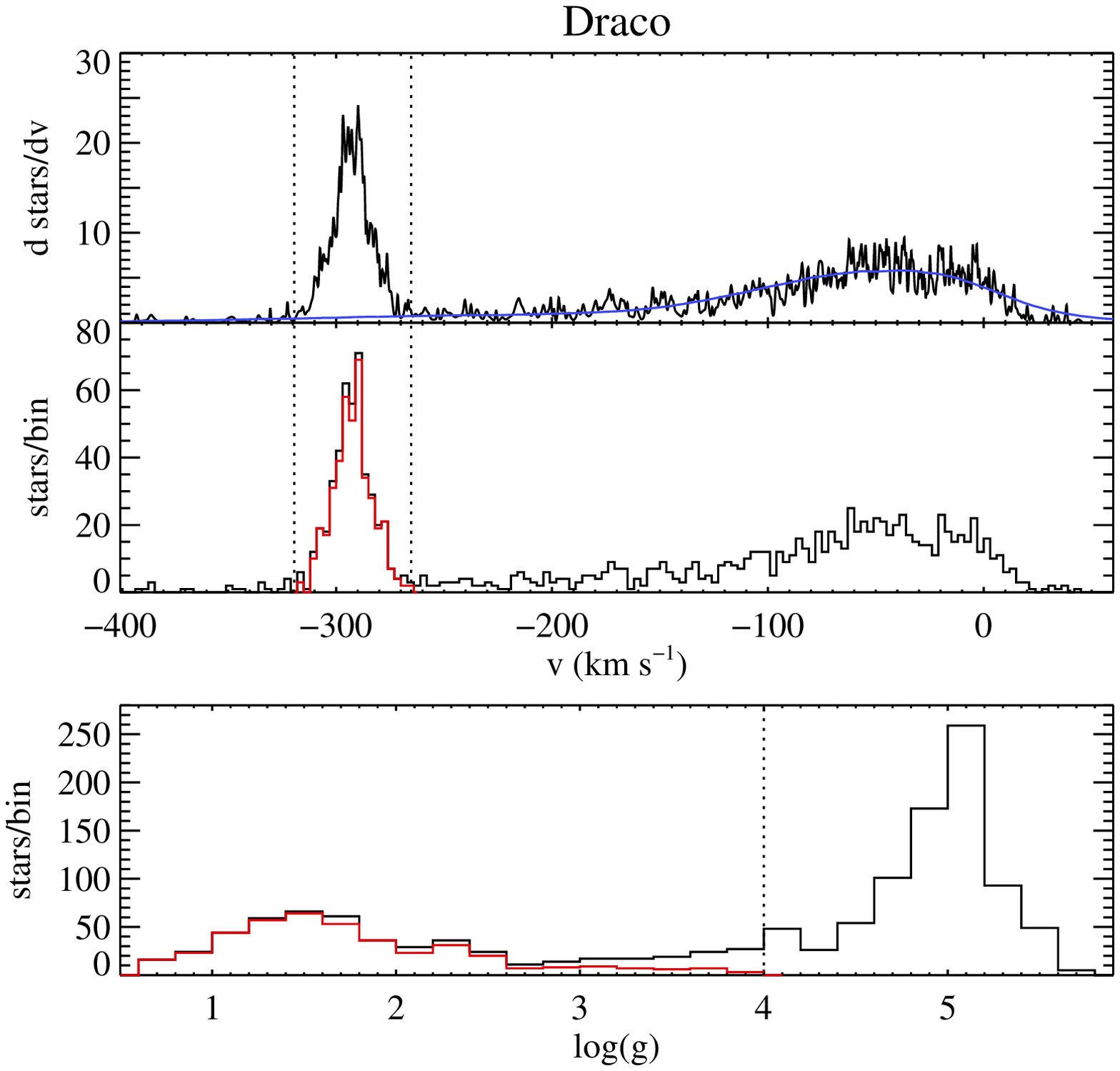}{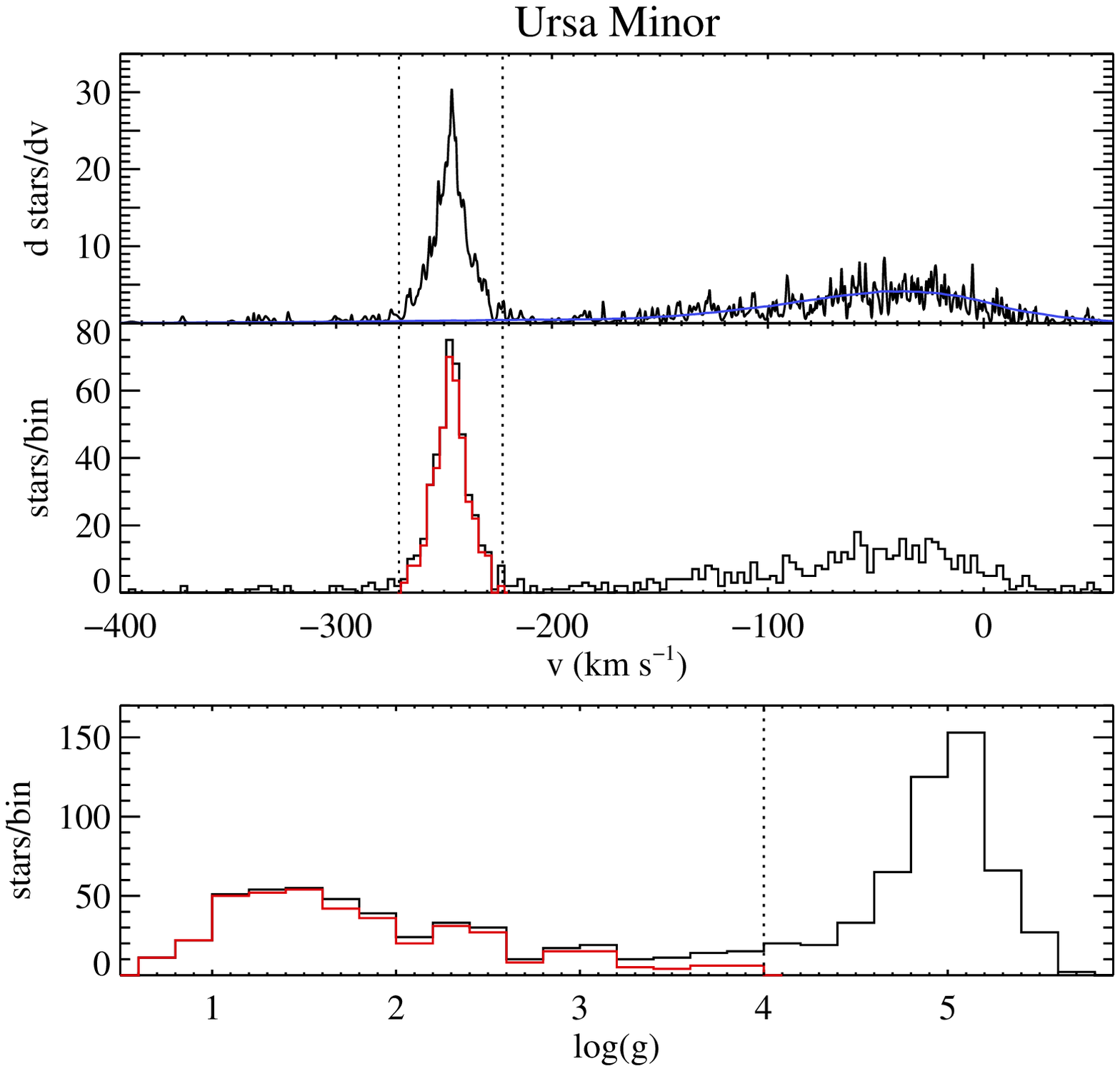}
\caption{Velocity and surface gravity histograms for Draco (left) and Ursa Minor (right). Top: Gaussian kernel density estimates of radial velocity from the MMT/Hectochelle data sets of \citetalias{walker2015} and Table \ref{tab:umi_table1} (black), and from a Besan\c{c}on model of MW foreground stars (blue). Middle: histogram of the average radial velocities in the MMT/Hectochelle data. The red line is the histogram of the final member selection. Vertical dotted lines demarcate the velocity boundary for membership. Bottom: histogram of the surface gravities in the MMT/Hectochelle data. The red line shows the histogram for the final member selection. The vertical doted line shows the membership cut in surface gravity.}
\label{fig_hist}
\end{figure*}

The samples of \citetalias{walker2015} and Table \ref{tab:umi_table1} contained larger data sets and probed fainter stars, so they incurred many more nonmembers than the other studies. For this reason, we spent extra care separating the members from the nonmembers by considering velocities and surface gravities. Figure \ref{fig_hist} serves as a visual aid for cuts that we made.

The following describes the procedure for defining the membership criteria for each galaxy using the MMT data of \citepalias{walker2015} and Table \ref{tab:umi_table1}. We started by taking the average radial velocity for each star. In the top panels of Figure \ref{fig_hist}, we plot Gaussian kernel density estimates of the radial velocities in black. This was done by adding together for each star a Gaussian with area equal to unity, location equal to the average radial velocity ($v=\Sigma\frac{v_i}{\sigma^2_i}/\Sigma\frac{1}{\sigma^2_i}$), and width equal to the weighted velocity uncertainty ($\sigma=(\Sigma\frac{1}{\sigma^2_i})^{-1/2}$). There is a sharp peak around -290 \kms{} for Draco and at -250 \kms{} for Ursa Minor, and a wide bump of Milky Way foreground stars that have slower radial velocities. 

For illustrative purposes, we generated a Besan\c{c}on model \citep{robin2003} that simulates the kinematics of Milky Way stars along the line of sight to each of the dwarfs. We made another Gaussian kernel density estimate for the Besan\c{c}on stars and set the width of each kernel equal to the median weighted velocity error of the dwarf under consideration. This value was 0.6 \kms{} for both dwarfs. We normalized the Besan\c{c}on model such that the area under the curve in the range of $-200<v<-20$ \kms{} for Draco (or $-160<v<-20$ \kms{} for Ursa Minor) was equal to the area under the black line over the same range. The model is shown as a blue line in Figure \ref{fig_hist}. 

In the middle panels of Figure \ref{fig_hist} we also plot regular histograms of the radial velocities. It is clear from both of these representations that there will be contamination from the Milky Way. The Besan\c{c}on models confirm the expectation that for such faint stars, most of the MW contaminants are main sequence stars. Therefore, contaminants will have significantly higher surface gravities, as can be seen in the bottom panels of Figure \ref{fig_hist}. The samples of \citetalias{walker2015} and Table \ref{tab:umi_table1} include measurements of surface gravity, so we have adopted a cutoff at $\log g=4.0$ to separate nonmember main sequence stars ($\log g \ge 4.0$) from possible Draco/Ursa minor member red giant and sub-giant stars ($\log g < 4.0$).

We then determined the radial velocity cuts that we should use by simultaneously deriving the systemic velocity and velocity dispersion of the dwarfs from the culled samples. The velocity dispersion and systemic velocity are found by a method of maximum likelihood described in \citet{walker2006}, which assumes that velocities are drawn from a Gaussian distribution. A first guess for the velocity membership boundaries is required for this method, so we used three times the standard deviation of a best-fit Gaussian, as shown in the middle panels of Figure \ref{fig_hist}. Stars within this velocity range were used to calculate the velocity dispersion and systemic velocity. The resulting velocity dispersion can be used to make new 3$\sigma$ velocity boundaries. Then we repeated the process of calculating the velocity dispersion and systemic velocity until it converged on an answer within 0.1 \kms{}. This took three to four iterations.

The resulting membership criteria for Draco were $-319.4<v<-265.2$ \kms{} and $\log(g) < 4.0$. The systemic velocity for Draco is $-292.3\pm0.4$ \kms{} with a velocity dispersion of $9.0\pm0.3$ \kms{}. Previously reported average, median, or systemic velocities are $-293.3\pm1.0$ \kms \citepalias{armandroff1995},  $-293.8^{+2.6}_{-2.7}$ \kms{} \citep{hargreaves1996b}, and $-290.7^{+1.2}_{-0.6}$ \kms{} \citepalias{wilkinson2004}, all of which agree with our findings. The velocity dispersion has been reported as $10.7\pm0.9$ or $8.5\pm0.7$, depending on the inclusion of one peculiar star \citepalias{armandroff1995}, $8.2\pm1.3$ \kms{} \citepalias{olszewski1995}, $10.5^{+2.2}_{-1.7}$ \kms{} \citep{hargreaves1996b}, and $9.1\pm1.2$ \kms{} \citep{mcconnachie2012}. Our measurements are in good agreement with the literature, especially given the large range of reported values.

\begin{figure*}
\epsscale{1}
\plottwo{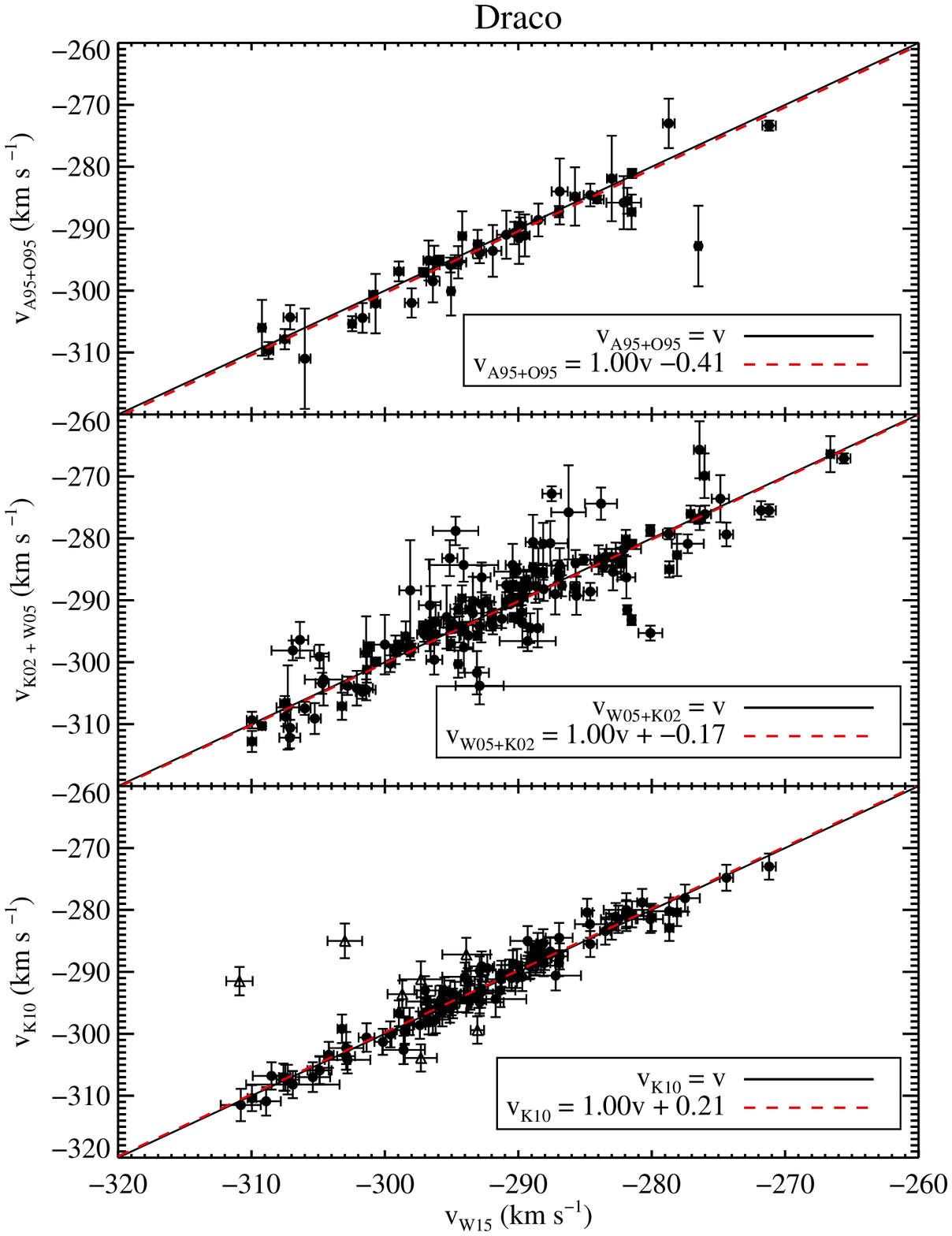}{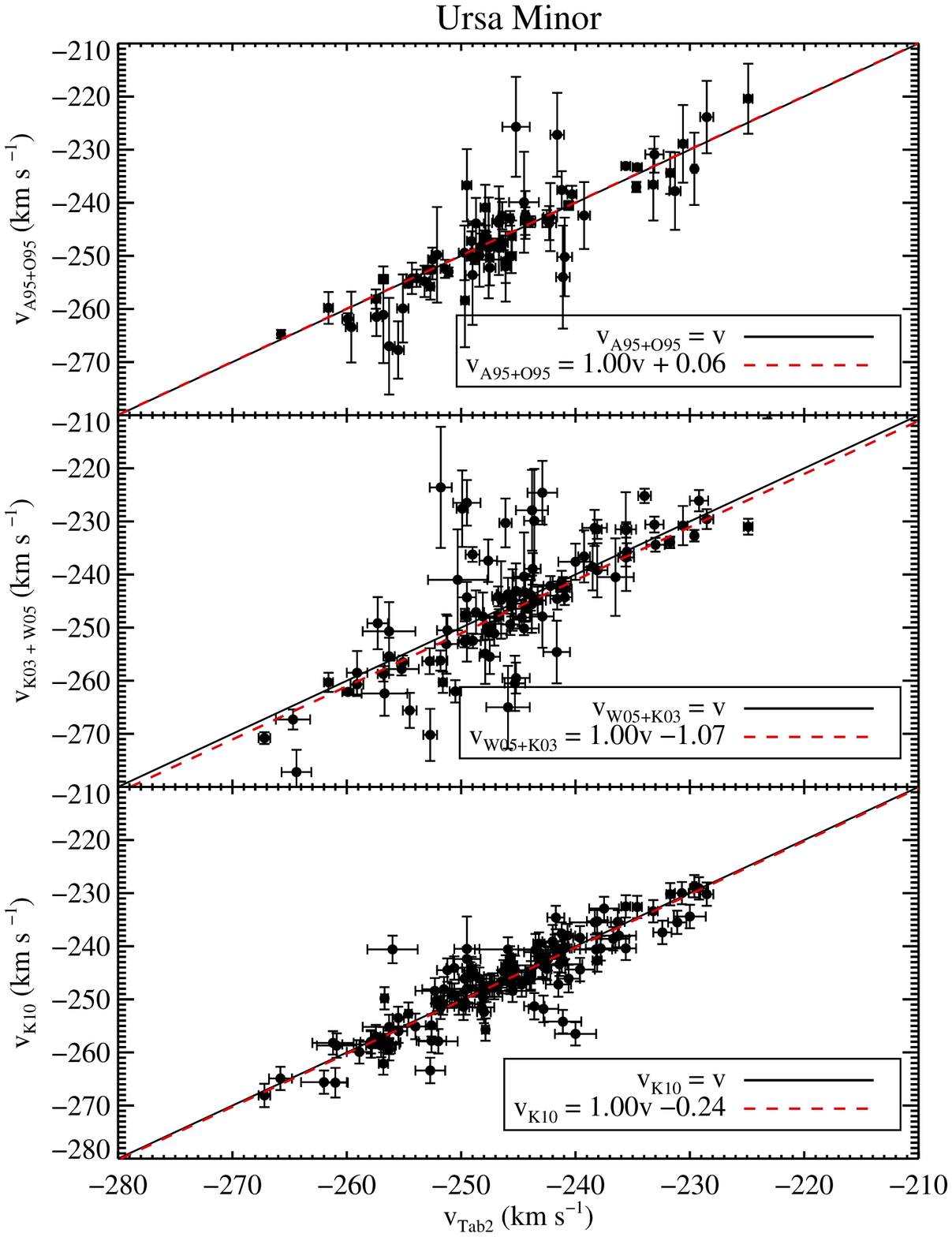}
\caption{Comparison of literature velocities to those of \citetalias{walker2015} for Draco (left) and Table \ref{tab:umi_table1} for Ursa Minor (right). The solid black line is where stars should fall if they have perfect agreement across studies, and the red dashed line shows the best-fit line with slope set equal to 1. The offset that is applied to each data set is the \textit{y}-intercept of the red dashed line.}
\label{fig_compare_vels}
\end{figure*}

For Ursa Minor, the membership criteria were $-270.9<v<-222.9$ \kms{} and  $\log(g) <4.0$. The systemic velocity for Ursa Minor is $-246.9\pm0.4$ \kms{} with a velocity dispersion of $8.0\pm0.3$ \kms{}. The average velocity was previously found to be $-247.2\pm1.0$ \kms{} \citepalias{armandroff1995}, $-249.2\pm1.5$ \kms{} \citep{hargreaves1994b}, and $-245.2^{+1.0}_{-0.6}$ \kms{} \citepalias{wilkinson2004}. Our measurement falls between these values. The velocity dispersion was listed as $10.4\pm0.9$ or $8.8\pm0.8$ \kms{}, depending on the inclusion of peculiar stars \citepalias{armandroff1995}, $10.5\pm2.0$ \kms{} \citepalias{olszewski1995}, $8.8$ \kms{} \citepalias{kleyna2003}, and $9.5\pm1.2$ \kms{} \citep{mcconnachie2012}. Our velocity dispersion is lower than all the other measurements, but the agreement is still within 1.3 $\sigma$ for all but one case. Several studies have found that the kinematics of Ursa Minor are better fit by a two-component model \citep{kleyna2003,wilkinson2004}. We do not explore more complicated dynamical models because the velocity distribution on the right side of Figure \ref{fig_hist} appears to be sufficiently Gaussian, and thus the method we used to determine membership should be valid. 

In the limiting case where all stars are binaries, we estimate that an average of only 1\% of member stars will not meet the velocity membership criteria due to binary orbital motion, and therefore the cuts we impose should not have an affect on our binary analysis.

\subsection{Correcting Systematic Offsets}\label{sec_velstandard}

Because we have incorporated data from a variety of different sources, it is possible that there are systematic offsets between the data sets. We have chosen \citetalias{walker2015} as the reference to which all the distributions will be shifted for Draco, and the data set presented in Table \ref{tab:umi_table1} as the reference for Ursa Minor. In Figure \ref{fig_compare_vels} we plot velocities from \citetalias{walker2015} or Table \ref{tab:umi_table1} along the \textit{x}-axis and velocities from the other studies along the \textit{y}-axis when stars exist in both catalogs. The black solid line is where stars would fall if they had perfect mean agreement. The red dashed line indicates the best fitting line with a slope set equal to 1. The \textit{y}-intercept of the line indicates the offset between \citetalias{walker2015} or Table \ref{tab:umi_table1} and others. We correct the velocities such that $v_{\mathrm{study\_corrected}} = v_{\mathrm{study}}-v_{\mathrm{offset}}$. Two outlier stars in \citetalias{kirby2010} were not used in the fit for Draco, and they are shown as open triangles in the bottom left panel of Figure \ref{fig_compare_vels}. 

\citetalias{armandroff1995} found that an offset of 1.59 \kms{} existed between their data and that of \citetalias{olszewski1995}. We add the same offset to the \citetalias{armandroff1995} data and plot the combined data set in the top panel to perform a comparison with \citetalias{walker2015} and Table \ref{tab:umi_table1}. This was necessary because only a couple stars were observed in common between the \citetalias{walker2015}/Table \ref{tab:umi_table1} and \citetalias{olszewski1995} data, making it impossible to perform the necessary comparison otherwise. 

\citetalias{kleyna2002}, \citetalias{kleyna2003}, and \citetalias{wilkinson2004} are also included in the same panel because the methods of observation and velocity extraction were identical, and they showed no signs of zero-point offsets between observing runs. 

In Draco the offsets are -0.17 \kms{} for \citetalias{kleyna2002} and \citetalias{wilkinson2004}, -0.41 \kms{} for \citetalias{olszewski1995} and \citetalias{armandroff1995}, and 0.21 \kms{}for \citetalias{kirby2010}. In Ursa Minor the offsets are -1.07 \kms{} for  \citetalias{kleyna2003} and \citetalias{wilkinson2004}, 0.06 \kms{} for \citetalias{olszewski1995} and \citetalias{armandroff1995}, and -0.24 \kms{} for \citetalias{kirby2010}.

\begin{figure*}
\centering
\plottwo{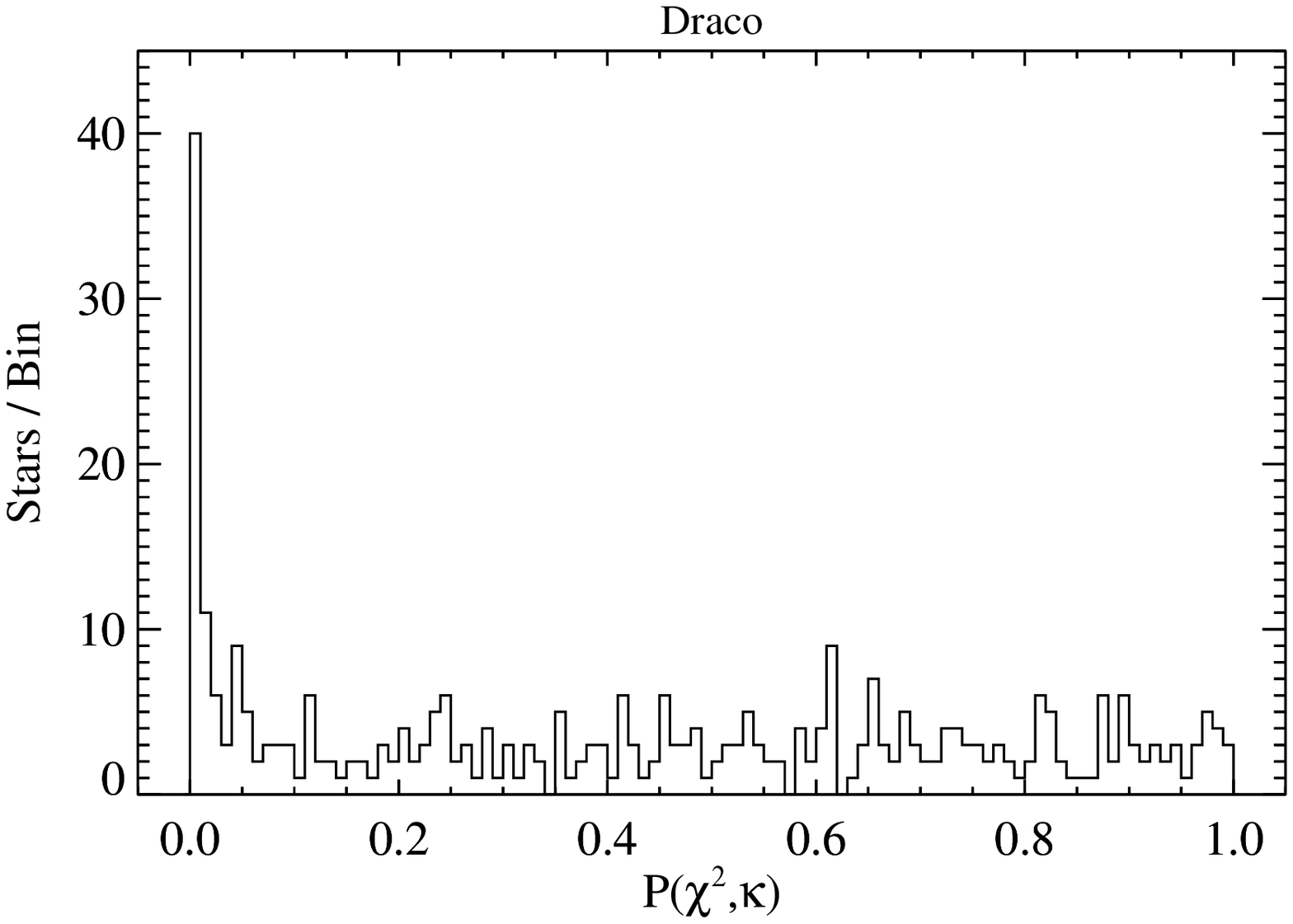}{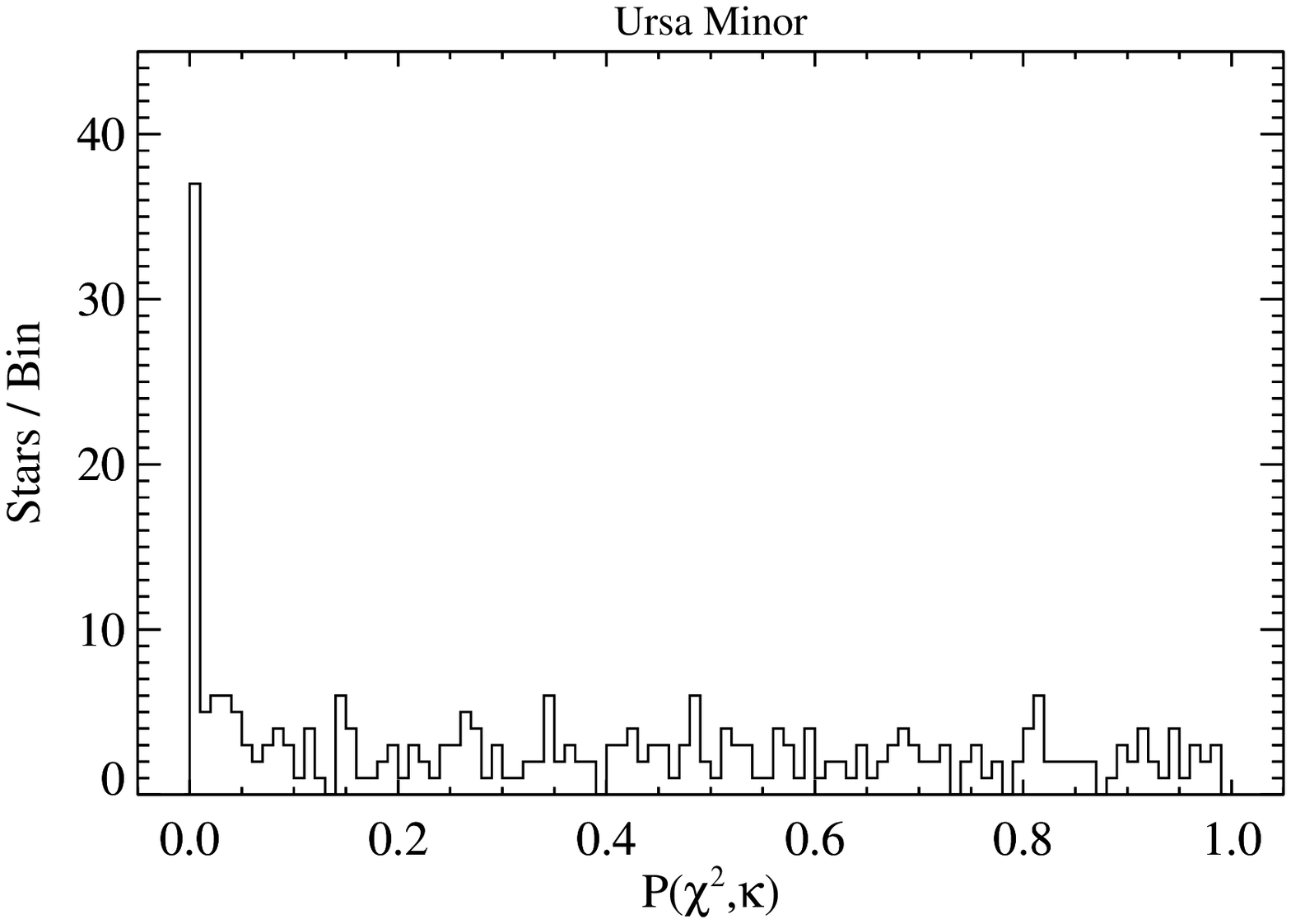}
\caption{Probability of exceeding $\chi^2$. The left panel is Draco and the right panel is Ursa Minor. With the exception of the lowest bin, which contains binaries, the histograms are uniform, suggesting that the velocity errors are properly reported.}
\label{fig_chisqu}
\end{figure*}

\begin{deluxetable*}{c c c c c r c}
\centering
\tablewidth{0pt}
\tablecaption{Velocities of RGB stars in Draco\label{tex_table_draco_vels}}
\tabletypesize{\scriptsize}
\tablehead{\colhead{Star ID} & \colhead{n} & \colhead{$\alpha_{\mathrm{J2000}}$} & \colhead{$\delta_{\mathrm{J2000}}$} & \colhead{HJD} & \colhead{$v$\tablenotemark{a}} & \colhead{Ref.} \\
 & & \colhead{(hh:mm:ss.ss)} & \colhead{(dd:mm:ss.ss)} & \colhead{(days)} & \colhead{(\kms{})} & }
 \startdata
Draco-001 &  2 & 17:15:36.04 & 57:48:34.40 & 2455707.8 & -287.1$\pm$ 1.4 & W15 \\
Draco-001 &  2 & 17:15:36.04 & 57:48:34.40 & 2455712.8 & -289.2$\pm$ 1.4 & W15 \\
Draco-002 &  2 & 17:15:41.94 & 57:37:05.50 & 2455707.8 & -299.5$\pm$ 2.2 & W15 \\
Draco-002 &  2 & 17:15:41.94 & 57:37:05.50 & 2455712.8 & -299.0$\pm$ 1.7 & W15 \\
Draco-003 &  2 & 17:16:01.59 & 57:59:19.10 & 2455707.8 & -282.9$\pm$ 0.8 & W15 \\
Draco-003 &  2 & 17:16:01.59 & 57:59:19.10 & 2452813.0 & -285.2$\pm$ 3.0 & W04 \\
\enddata
\tablenotetext{a}{Velocities after correcting for systematic offsets. Only stars with multi-epoch velocity measurements are included.}
\tablecomments{This table is published in its entirety in the electronic edition of the Astrophysical Journal. A portion is shown here for guidance regarding its form and content.}
\end{deluxetable*}

\begin{deluxetable*}{c c c c c r c}
\centering
\tablewidth{0pt}
\tablecaption{Velocities of RGB stars in Ursa Minor\label{tex_table_ursaminor_vels}}
\tabletypesize{\scriptsize}
\tablehead{\colhead{Star ID} & \colhead{n} & \colhead{$\alpha_{\mathrm{J2000}}$} & \colhead{$\delta_{\mathrm{J2000}}$} & \colhead{HJD} & \colhead{$v$\tablenotemark{a}} & \colhead{Ref.} \\
 & & \colhead{(hh:mm:ss.ss)} & \colhead{(dd:mm:ss.ss)} & \colhead{(days)} & \colhead{(\kms{})} & }
 \startdata
UrsaMinor-001 &  2 & 15:04:55.74 & 66:28:39.91 & 2454616.8 & -235.5$\pm$ 0.8 & Tab2 \\
UrsaMinor-001 &  2 & 15:04:55.74 & 66:28:39.91 & 2455232.9 & -234.9$\pm$ 1.3 & Tab2 \\
UrsaMinor-002 &  2 & 15:05:29.84 & 67:12:43.69 & 2455659.7 & -245.9$\pm$ 1.9 & Tab2 \\
UrsaMinor-002 &  2 & 15:05:29.92 & 67:12:43.52 & 2452769.0 & -263.9$\pm$ 7.8 & W04 \\
UrsaMinor-003 &  2 & 15:05:44.64 & 67:03:11.11 & 2454615.3 & -250.5$\pm$ 0.5 & Tab2 \\
UrsaMinor-003 &  2 & 15:05:44.70 & 67:03:11.12 & 2452769.0 & -260.9$\pm$ 2.1 & W04 \\
\enddata
\tablenotetext{a}{Velocities after correcting for systematic offsets. Only stars with multi-epoch velocity measurements are included.}
\tablecomments{This table is published in its entirety in the electronic edition of the Astrophysical Journal. A portion is shown here for guidance regarding its form and content.}
\end{deluxetable*}

\subsection{Velocity Uncertainty}\label{sec_velerrors}

A crucial element of this analysis is having accurate radial velocity errors. Underestimated errors will inflate the binary fraction, while overestimated errors will decrease it. To determine if the errors are an accurate representation of the scatter in the velocity data, we use the $\chi^2_\kappa$ statistic, defined as 
\begin{equation}
\chi^{2}_{\kappa} = \frac{1}{\kappa} \sum_{i}^{n}\Big(\frac{v_{i}-\langle v\rangle}{\sigma_{i}}\Big)^{2} ,
\label{eq_chisqu_draumi}
\end{equation}
where $v_i$ and $\sigma_i$ are a single velocity and corresponding error measurement, $\langle v\rangle$ is the average velocity for a star, $\kappa = n-1$ is the number of degrees of freedom, and $n$ is the number of observations per star. The probability of exceeding $\chi^2_\kappa$ is $P(\chi^2,\kappa)$. In Figure \ref{fig_chisqu} we plot histograms of $P(\chi^2,\kappa)$ under the assumption that all stars are velocity non-variables. If the errors are accurate and there are no intrinsic velocity variables with resolvable $\Delta v$'s, then the distribution should be flat. If the errors are over/underestimated and there are no stars with resolved $\Delta v$'s, the histograms would be biased toward higher/lower P values. If binaries exist with $\Delta v$'s comparable to the observational errors and the errors are well-determined, then the lowest bin, $0<P(\chi^2,\kappa)<0.01$, would be enhanced relative to the mean $\chi^2$ value. 

As a test of the precision of our error estimates and the existence of binaries, we have fit two lines to the histograms in Figure \ref{fig_chisqu}; one line has a fixed slope of zero and the other has a variable non-zero slope. In Draco, the flat line had a \textit{y}-intercept of $2.96\pm0.15$, and the expectation was 3.03 (calculated as the number of stars with $P(\chi^2,\kappa)>0.01$ divided by 99 bins). The slope in the second line had a 1 $\sigma$ error bar as large as the value, indicating that it is consistent with being flat. The good agreement of the line being flat indicates that the velocity errors in Draco accurately represent the data. In Ursa Minor, the results are very similar. The flat line had a \textit{y}-intercept of $2.47\pm0.14$, and the expectation was 2.49. Once again, the slope of the second line had errors as large as the value. We draw the same conclusion for Ursa Minor as Draco: the velocity errors are not over- or underestimated. 

\begin{figure*}
\epsscale{0.9}
\plottwo{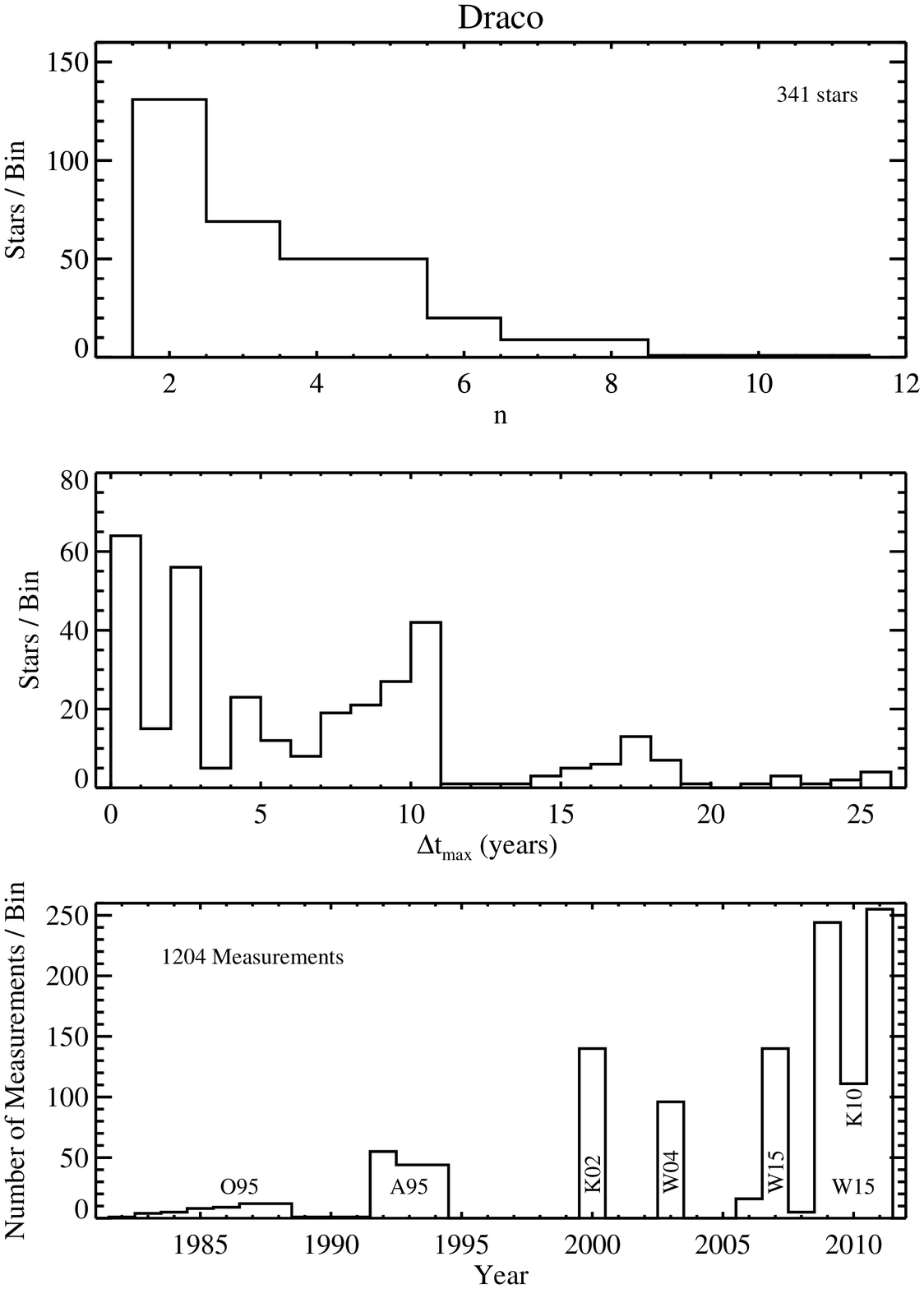}{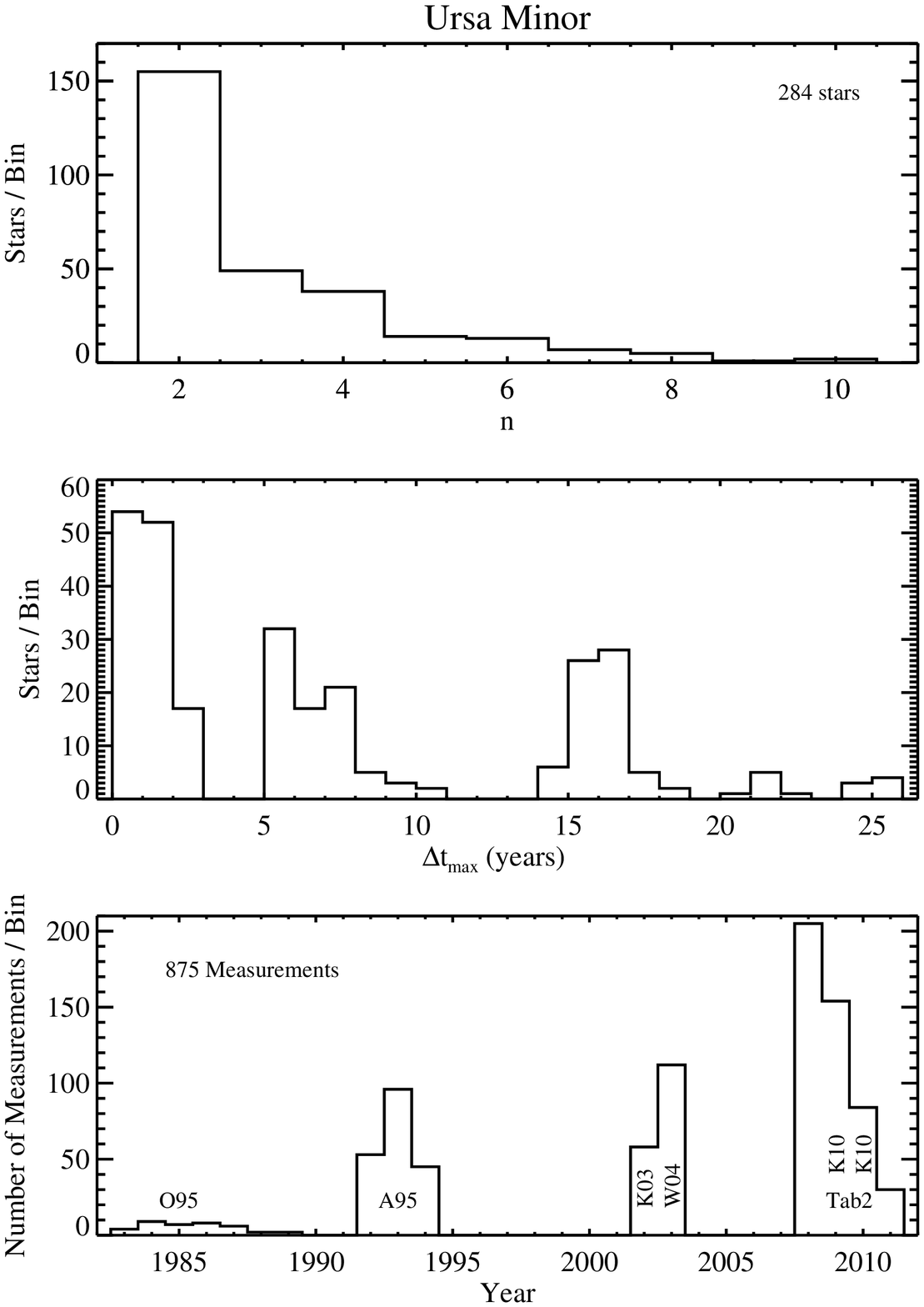}
\caption{Summary of Draco data (left) and Ursa Minor (right). Top: histogram of number of observations per star. Middle: histogram of maximum time interval length per star. Bottom: histogram of number of measurements taken per year. Bins are labeled with the paper that produced the measurements.}
\label{fig_deltat}
\end{figure*}

\subsection{Summary of Velocity Data}

In Draco there are 692 unique member stars, 341 of which have multiple observations. There are a total of 1204 velocity measurements for the subset of stars with multiple observations. These data are listed in Table \ref{tex_table_draco_vels}. In Ursa Minor there are 680 unique member stars. A total of 284 of them have multiple observations. There are a total of 875 measurements for the stars with multiple observations, which are listed in Table \ref{tex_table_ursaminor_vels}. In both of these tables, column 1 lists the identifier that we assign to the star; column 2 lists the number of observations; column 3 lists the right ascension; column 4 lists the declination; column 5 lists the heliocentric Julian date; column 6 lists the radial velocity and error; and column 7 lists the paper from which the velocity measurement originated. Because we applied offsets to most of the velocity data to put them on the same standard (see Section \ref{sec_velstandard}), the velocities we report in the tables will not match the values listed in the original papers.

Some important aspects of these velocity data are highlighted in Figure \ref{fig_deltat}. In the top panel we plot a histogram of the number of observations per star, $n$. The maximum number of observations in Draco is 11, and in Ursa Minor the maximum is 10. The middle panel is a histogram of the amount of time elapsed between the first observation and the last observation for each star. Both dwarfs have a handful of stars with time intervals as long as 25 years. Finally, the bottom panel is a histogram of the number of measurements taken per year. The bins are labeled with the study that produced the measurements.

\section{Methodology}\label{sec_methods}

The term ``binary fraction'' has taken on several slightly different definitions and names, such as companion frequency, multiplicity frequency, and multiplicity rate \citep[e.g., ][]{olszewski1996,duchene2013}. In the present study, we consider two stars that are gravitationally bound to one another to be a binary system. We define the binary fraction, $f$, to be the fraction of all apparently single stars that turn out to be binary systems based, in our case, on their velocity variability. We do not consider photometric binaries because the remoteness of the systems makes these hard to detect, though wide binaries may exist in dSphs \citep[e.g.,][]{penarrubia2016}. The constituent stars of a binary system do not get double counted by this definition of the binary fraction. This definition is sufficient for our study because we are considering only binary systems containing red giants, which are unlikely to pair with similar stars due to their comparatively short lifetimes.

The goal of this study is to determine the binary fractions of the stellar populations comprising the Draco and Ursa Minor dSph galaxies. Our analysis considers all of the stars as a collection and does not distinguish which stars are most likely to be binaries. The method for determining the binary fraction that we adopt in this chapter is similar to that described in \citet[][hereafter Paper I]{spencer2017b}, but with some changes. The primary steps of the process involve defining the binary parameter distributions (Section \ref{sec_binparams}), running Monte Carlo (MC) simulations of the velocity variability (Section \ref{sec_mcsims}), performing a Bayesian analysis on the data and simulations (Section \ref{sec_bayesian}), and extracting a binary fraction from the posterior (Section \ref{sec_posterior}).

\subsection{Binary Parameters}\label{sec_binparams}

\begin{figure}
\epsscale{1.1}
\plotone{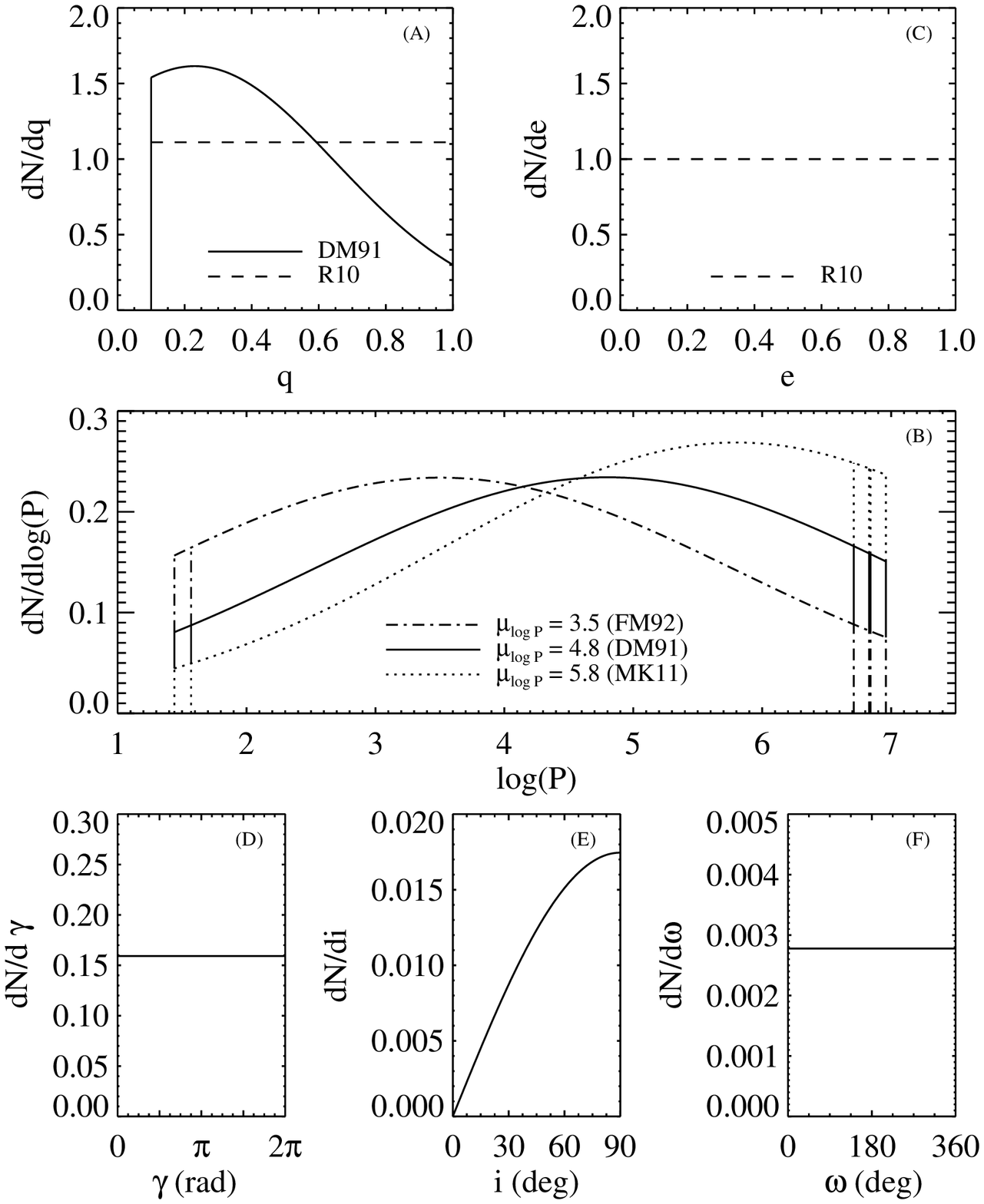}
\caption{Binary parameter distributions used in our simulations. Panels (A)--(C) are based on observations or theory from \citet[][solid lines]{duquennoy1991}, \citet[][dashed lines]{raghavan2010}, \citet[][dotted line]{marks2011}, and \citet[][dot-dashed line]{fischer1992}. Vertical lines in Panel (B) indicate the range in the upper and lower boundaries caused by the mass ratio. Panels (D)--(F) are based on observational geometry.}
\label{fig_bin_params}
\end{figure}

As described in \citetalias{spencer2017b}, there are seven parameters that go into determining the orbital radial velocity of a binary. They are mass of the primary ($m_{1}$), mass ratio ($q$), period ($P$), eccentricity ($e$), true anomaly ($\theta$), inclination ($i$), and argument of periastron ($\omega$). The first four parameters are intrinsic to the system, and the last three reflect the geometry of the system with respect to the observer. The equation that relates these parameters to the orbital radial velocity is
\begin{equation}
v_{r,orb} = \frac{q \sin i}{\sqrt{1-e^{2}}}\bigg(\frac{2\pi G m_{1}}{P(1+q)^{2}}\bigg)^{1/3}\big(\cos(\theta+\omega)+e\cos\omega\big) .
\label{eq_orv}
\end{equation}
For a derivation and additional details of this equation, see \citet{green1985} or \citet{spencer2017c}.

\subsubsection{Mass of Primary, $m_1$}

The mass of the primary, $m_1$, can be set at a fixed value of $m_1=0.8$ M$_{\odot}$ because all of the stars in our sample are extremely old and are located along the red giant branch. All of the other parameters will vary from binary to binary. Mass ratio, period, and eccentricity are somewhat dependent on one another, so they will be drawn in the same order every time to ensure that the dependencies are preserved.

\subsubsection{Mass Ratio, $q$}

In this study, mass ratio is defined as $q=m_2/m_1$, where $m_1$ is the primary red giant star and $m_2$ is the secondary star. We assume that the secondary is a non-giant. We select the minimum mass ratio to be $q_{\mathrm{min}}=0.1$. This requires the secondary companion to be a hydrogen-burning star with a mass of at least 0.08 M$_{\odot}$. For these stellar populations, the main sequence turnoff mass is 0.8 M$_{\odot}$, which corresponds to a maximum value of $q_{\mathrm{max}}=1$.

We selected two mass ratio distributions to investigate in our simulations. The first is a normal distribution from \citet[][hereafter DM91]{duquennoy1991}, which is described by
\begin{equation}
\frac{dN}{dq} \propto \exp(-\frac{(q-\mu_{q})^{2}}{2\sigma_{q}^{2}}).
\label{eq_massratio1}
\end{equation}
The parameters that they found to best describe their data were $\sigma_q=0.42$ and $\mu_q=0.23$. 
The second distribution,
\begin{equation}
\frac{dN}{dq} = const, 
\label{eq_massratio2}
\end{equation}
is constant across all mass ratios. This was used in \citet[][hereafter R10]{raghavan2010}, as well as many other papers. The two distributions are plotted in Panel (A) of Figure \ref{fig_bin_params}.

\subsubsection{Period, $P$}

The period distribution has the largest effect on the binary fraction \citep{minor2013,spencer2017b}. For this reason, we consider three different period distributions to get a better understanding of the range of allowable binary fractions. This also allows us to rule out very high or very low binary fractions. Most studies have found that a log-normal form provides the best fit to the observed periods of binary stars in the solar neighborhood, so we will select this form and change the parameters in the equation. The notation for this period distribution is
\begin{equation}
\frac{dN}{d\log P} \propto \exp\bigg(-\frac{(\log P - \mu_{\log P})^{2}}{2\sigma_{\log P}^{2}}\bigg),
\label{eq_period1}
\end{equation}
where $\mu_{\log P}$ is the center of the distribution and $\sigma_{\log P}$ is the width. The period is expressed in units of days. In addition to fixing the functional form, we also fixed the second parameter, $\sigma_{\log P}$, at 2.3 log(days) \citepalias{duquennoy1991}. This makes it easier to discern the effect that $\mu_{\log P}$ has on the inferred binary fraction.
We first select $\mu_{\log P}=4.8$. This value was found by \citetalias{duquennoy1991} to provide the best fit to F7 to G9 type stars in the Solar Neighborhood. A larger study by \citetalias{raghavan2010} recovered $\mu_{\log P}=5.03$ from their sample of F6 to G2 type stars. Because these two values are so similar, we opted to only use the first prescription. 

The second value of $\mu_{\log P}$ we considered is for K and M-dwarf stars \citep[][hereafter FM92]{fischer1992}. They found that the peak in $\log P$ occurred at much lower values between 3.5 and 4.9. We selected the smaller value for $\mu_{\log P}$ because this provided a distribution that was more discrepant from the one defined previously.

The third period distribution is theoretical in nature and comes from \citet[][hereafter MK11]{marks2011}. They explored how the shape of the period distribution for main sequence stars might change with respect to spectral type, birth cluster clump size, and star formation rate. We selected their solution for a dwarf irregular galaxy in their Figure 9. This distribution is not log-normal, but instead is skewed toward longer periods. We fit a log-normal to their distribution and find that $\mu_{\log P}=5.8$ does the best job of reproducing it, and so we adopt this for the last value of $\mu_{\log P}$. 

The minimum and maximum binary periods expected for red giants in Draco and Ursa Minor can be constrained by considering the semi-major axes, $a$, compatible with these stars in such environments. The minimum semi-major axis occurs when the stellar surfaces are just out of contact. The primary star is a red giant with a radius much larger than the secondary, so we estimate $a_{min}$ as the radius of the primary. Assuming a mass of 0.8 M$_{\odot}$ and a surface gravity of about 10 cm s$^{-2}$, the radius works out to be about 0.21 au. Using Kepler's third law, this corresponds to a period of $\log P_{min}=1.57$ for a mass ratio of 0.1 or $\log P_{min}=1.44$ for a mass ratio of 1.

The maximum semi-major axis is the maximum extent that a binary can reach before the gravitational force from its partner is less than that of neighboring stars in the galaxy. If we consider the gravitational unbinding of a binary due to the encounter with another star to be a ``collision,'' then the minimum semi-major axis can be thought of as the cross-section in the equation for mean free path. This yields the equation $a_{\mathrm{max}}=(\pi\sigma_v t \lambda)^{-1/2}$, where $\sigma_v$ is velocity dispersion, $t$ is the average age of the stars, and $\lambda$ is the number density of the stars. We calculate the number density by converting central luminosity density to mass density with the assumptions that a star has an average mass of 0.4 $M_\odot$ and that $L/L_\odot\propto(M/M_\odot)^4$. For Draco, we used 9.0 \kms{} as the velocity dispersion (Section \ref{sec_membership}), 10 Gyr as the average age \citep{aparicio2001}, and 0.008 L$_\odot$ pc$^{-3}$ as the central luminosity density \citep{mateo1998}. For Ursa Minor, we used 8.0 \kms{} as the velocity dispersion (Section \ref{sec_membership}), 10 Gyr as the average age \citep{carrera2002}, and 0.006 L$_\odot$ pc$^{-3}$ as the central luminosity density \citep{mateo1998}. This places $\log P_{max}$ between 6.71 and 6.84 log(days) for Draco, and between 6.83 and 6.96 log(days) for Ursa Minor, depending on the mass ratio. These three period distributions and the limits are shown graphically in Panel (B) of Figure \ref{fig_bin_params}.

\subsubsection{Eccentricity, $e$}

Eccentricity only has a small effect on the observed binary fraction \citep{minor2010,spencer2017b}; therefore we selected a single distribution for this parameter. We used the one from \citetalias{raghavan2010}:
\begin{equation}
\frac{dN}{de} \propto const.
\label{eq_eccenricity1}
\end{equation}
Another choice would have been a thermal distribution (i.e., $dN/de=2e$), but it has been shown by \citet{duchene2013} that binary main sequence stars with periods greater than 100 days do not follow this trend. We note that binaries with short periods (on the order of 10--20 days) will have circular orbits. However, we do not need to include this condition because binaries with such short periods would have been destroyed as the primary ascended the red giant branch \citep{iben1993,nie2012}. 

The eccentricity can range from 0 to 1, but in many cases the maximum value must be lower to prevent the stars from colliding, as can happen with certain combinations of $P$ and $q$. This limit is set by $e_{max}=1-(a_{min}/a)$, where $a$ is the solution for the semi-major axis from Kepler's third law, given $q$ and $P$, as noted previously.

\subsubsection{True Anomaly, $\theta$}

The true anomaly, $\theta$, is the angle between the lines connecting the periastron to the focus and the focus to the location of the star along its orbit. It dictates where the star is in its orbit. This angle is dependent on the eccentricity and period in such a way that it has no analytical solution. Instead, we define a distribution for the area swept out by the star since it passed periastron. Kepler's second law states that within a gravitationally bound binary system, the radius vector of each component will sweep out an equal area in a given amount of time relative to the position of the other star. By normalizing the area so that periastron corresponds to 0 or $2\pi$ and apastron corresponds to $\pi$, we can redefine area as the mean anomaly, $\gamma$. The frequency of a star being observed at any $\gamma$ is constant; thus 
\begin{equation}
\frac{dN}{d\gamma}=\mathrm{const.}
\label{eq_theta}
\end{equation}
The true anomaly can then be numerically solved for using the mean anomaly and the eccentricity. Once the mean anomaly at the time of the first observation is selected, the location of the star in its orbit at all later times will be described by $\gamma=\gamma_{1}+(2\pi\Delta t/P)$, where $\Delta t$ is the time elapsed since the first observation.

\subsubsection{Inclination, $i$}

Inclination, $i$, is the angle between the observer's line of sight and the normal to orbital plane of the system. It ranges from 0 (face on) to 90 degrees (edge on) and has the form
\begin{equation}
\frac{dN}{d i} \propto \sin(i).
\label{eq_inclination}
\end{equation}

\subsubsection{Argument of Periastron, $\omega$}

Finally, the argument of periastron, $\omega$, is the angle between the ascending node of the orbit and the periastron point. It ranges from 0$\degree$ to 360$\degree$. All values have an equal probability of occurring so we can write the distribution as
\begin{equation}
\frac{dN}{d\omega}=\mathrm{const.}
\label{eq_omega}
\end{equation}

The distributions of all the parameters described in this section are plotted in Figure \ref{fig_bin_params}.

\subsection{Monte Carlo Simulations}\label{sec_mcsims}

The purpose of the Monte Carlo (MC) simulations produced for this study is to generate a series of radial velocities that would be expected for a given binary fraction and compare those velocities with our observed radial velocities. The MC simulation that most resembles the data will tell us what the binary fraction is, under the assumptions of the simulations and given the binary parameter distributions described in Section \ref{sec_binparams}. We have chosen to perform our analysis on the entire set of velocity data simultaneously for a given dwarf galaxy---Draco or Ursa Minor---rather than considering the binarity of each individual star, as has been done by others \citep{minor2010,martinez2011,cottaar2014}. 

Since we are only concerned with velocity variability, the mean motion of each star within the potential of a dwarf galaxy is irrelevant for this part of the analysis  We consider the change in velocity by defining a statistic as
\begin{equation}
\beta = \frac{|v_{i}-v_{j}|}{\sqrt{\sigma^{2}_{i}+\sigma^{2}_{j}}},
\label{eq_beta}
\end{equation}
where $v$ is the radial velocity and $\sigma$ is the corresponding uncertainty in velocity. The subscripts $i$ and $j$ denote different observations of the same star. Stars with more observations will have more $\beta$'s and thus provide better leverage on the binary fraction. The number of $\beta$'s per star is calculated by $n(n-1)/2$, where $n$ is the number of observations per star. 

The collection of $\beta$'s for a dSph is what we aim to reproduce with MC simulations. Our simulations employ data from the observational catalogs of Draco and Ursa Minor---radial velocity uncertainty and the Julian date of each radial velocity measurement---along with the binary fraction, binary parameter distributions, and parameter limits described in Section \ref{sec_binparams}. With these data and parameter inputs, we carry out the following steps to generate MC simulations of $\beta$, using binary fraction, $f$, as the primary variable:

\begin{enumerate}
\item While going through each star in the input data sets for Draco or Ursa Minor, we randomly assign the star as a member of a binary system based on the binary fraction, $f$, being tested. 
\item If the star is a binary according to Step 1, then we randomly select a value for each of the seven binary parameters according to the distributions described by Equations \ref{eq_massratio1}--\ref{eq_omega}.  If the star is not a binary, this step is skipped.
\item We then calculate the radial velocity for the star. If the star is a binary, this value comes from Equation \ref{eq_orv}. If the star is not a binary, this value is 0 \kms. (Zero signifies that the star has no velocity variation induced by a binary.)
\item We then resample the velocity from Step 3 by adding a Gaussian deviate with standard deviation equal to the velocity uncertainty for that observation.
\item Steps 3-4 are repeated $n$ times, where $n$ is the number of observations for that star. All binary parameters from Step 2 are kept the same for an individual star except for the true anomaly, $\theta$. This parameter is advanced by an amount corresponding to $\Delta t$, as described in the paragraph surrounding Equation \ref{eq_theta}.
\item We then calculate the $\beta$'s for that star by using Equation \ref{eq_beta}.
\item Steps 1-6 are repeated for all stars in a given galaxy's database. (There were 341 stars in Draco and 284 in Ursa Minor.)
\item Steps 1-7 are repeated $\eta$ times. Unless noted, we typically adopted $\eta=10^4$.
\item If a range of binary fractions was being investigated, we then repeated Steps 1-8 for each binary fraction under consideration. In most cases, we tested binary fractions from 0 to 1 in increments of 0.01.
\end{enumerate}

\subsection{Bayesian Technique}\label{sec_bayesian}

Next we compare the ``base distribution''---the distribution of $\beta$ values from the observations, $\beta_{obs}$---with the distributions of $\beta$'s from the MC simulations, $\beta_{mod}$. Our aim is to determine the probability that the base distribution can be reproduced by $\beta_{mod}$, given a certain binary fraction, $f$. One way to address this is through a Bayesian analysis. To begin, we can write Bayes' Theorem as 
\begin{equation}
P(f|D,M) = \frac{P(D|f,M)P(f|M)}{P(D|M)}, 
\label{eq_bayes}
\end{equation}
where the data, $D$, is $\beta_{obs}$ from the observations, and the model, $M$, is $\beta_{mod}$ from the MC simulations. $P(f|M)$ is the prior and contains any previous knowledge that we might have had on the binary fraction before we began the analysis. This term is set equal to 1 because we have no prior constraints on $f$. The denominator, $P(D|M)$, is a normalization factor that we select such that the integral of the posterior, $P(f|D,M)$, is equal to unity. These two simplifications mean that the posterior is directly proportional to the likelihood of the data being produced by a given binary fraction and set of models, $P(D|f,M)$.

\begin{figure}
\epsscale{1}
\plotone{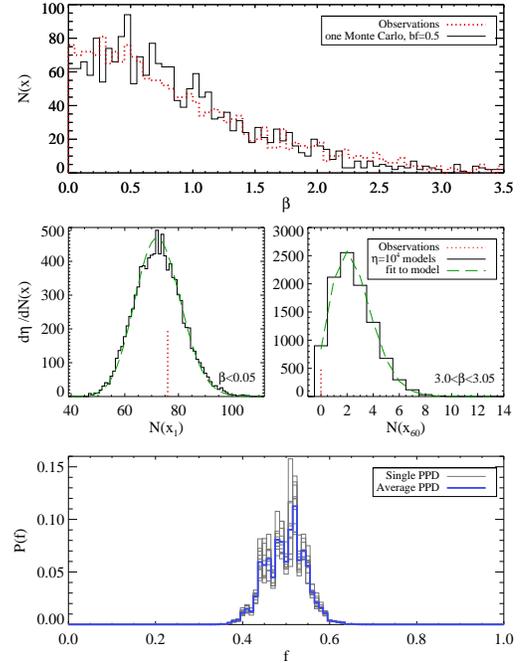}
\caption{Top: the histogram of $\beta_{obs}$ is shown as a red dotted line. For readability, we plot only one histogram of a single Monte Carlo simulation with $f=0.5$, which is shown by the black solid line. Middle: histograms of the number of $\beta_{mod}$ in bin 1 ($\beta<0.05$, left) and bin 60 ($3.0<\beta<3.05$, right). The $\eta=10^4$ MC models are fit with a green long-dashed line. The vertical red dashed line marks the number of $\beta_{obs}$ in the bin. Bottom: PPDs for a single bin size are shown in gray. The normalized sum of these is shown in blue. These particular PPDs are for Draco with a normal mass ratio distribution and a log-normal period distribution located at $\mu_{\log P}=4.8$.}
\label{fig_example}
\end{figure}

Deriving the equation for likelihood is somewhat complicated, so we include Figure \ref{fig_example} to help illustrate the process for the case of $f=0.5$. We start by placing the $\beta$'s into bins according to their value. In the top panel of Figure \ref{fig_example}, we show this ordering as a red dashed histogram for $\beta_{obs}$. A similar histogram for $\beta_{mod}$ is shown as a solid black line. We only show one of these to enhance readability, but there are in fact $\eta=10^4$ of these $\beta_{mod}$ histograms\footnote{The ability of the MC simulations to reproduce $\beta_{obs}$ is discussed in Section \ref{sec_results}. Cumulative histograms of $\beta$ are plotted later in Figure \ref{fig_ref}, which show how much $\beta$ can vary between MC simulations. Histograms of MC simulations for $f=0$ are also in Figure \ref{fig_ref}. Small values of $\beta$ are dominated by measurement error, whereas large values, greater than about 3--4, are most likely caused by binary motion.}. We then define the number of $\beta_{obs}$ in bin $x$ as $N(x)_{obs}$. This is represented by the vertical red dashed line in the second panel of Figure \ref{fig_example} for the first bin. The number of $\beta_{mod}$ in a given bin, $x$, for a certain simulation number, $j=\{0,1,2...\eta\}$, is $N(x|j)_{mod}$. The histogram for $N(x_1|j)_{mod}$ is shown as a black solid line in the second panel of Figure \ref{fig_example}.

One could, in principle, use the histogram of $N(x=1|j)_{mod}$ as the probability mass function to compute a likelihood. However, this would yield a very noisy posterior. Instead, it is best to use a smooth function for the probability mass function. We have found that a Poisson function does a good job of reproducing $N(x|j)_{mod}$, which can be written as
\begin{equation}
\phi(N(x)|\mu_x)=\frac{\mu_x^{N(x)} \exp(-\mu_x)}{N(x)!}.
\label{eq_poisson}
\end{equation}
There is only one parameter in this distribution, $\mu_x$, the average number of $\beta_{mod}$ in the bin $x$  ($\mu_x=(1/\eta)\sum_{j=1}^{\eta}N(x|j)_{mod}$). In cases where $\mu_x$ was greater than 100, the equation became numerically unstable, so we approximated it as a Gaussian with location $\mu_x$ and standard deviation equal to $\sqrt{\mu_x}$. We plot the corresponding Poisson function for the first bin, $\phi(N(x_1)|\mu_1)$, as a green long-dashed line in the middle left panel of Figure \ref{fig_example}. Bin number 60, $\phi(N(x_{60})|\mu_{60})$, is also plotted on the right to show that a Poisson function does a good job of representing $N(x|j)_{mod}$ for bins with either large or small values of $\beta$.

The likelihood for a single bin is then Equation \ref{eq_poisson} evaluated at $N(x)_{obs}$. This can easily be extrapolated to the likelihood over all bins by taking the product of the likelihoods from each bin. Recalling that the posterior is proportional to the likelihood, we finally arrive at 
\begin{equation}
P(f|D,M) \propto \prod_{x}\phi(N(x)_{obs}|\mu_x).
\end{equation}
We repeated this calculation over all $f$ and normalized it such that $\sum_{f=0}^{f=1}P(f|D,M)=1$.

There are two key parameters that we have yet to discuss that play a role in the posterior. They are bin size and number of bins. The bin size must be smaller than the largest $\beta$ for a given $f$. This limit is set by the $f=0$ case and works out to be about 2.5. If the bin size is larger than this value, then the probabilities for $f=0$ and other small $f$ will be indistinguishable. In addition, $N(x|j)_{mod}$ is only well fit by a Poisson when the bin size is $\lesssim 0.05$. Bin sizes larger than 0.05 but smaller than 2.5 can still recover the binary fraction, but a skewed normal must be used in place of a Poisson \citepalias{spencer2017b}. 

The number of bins must be large enough to encapsulate all of the $\beta$'s, both observed and modeled. The largest values of $\beta_{mod}$ are usually around 90, which equates to about 2000 bins. Additional bins that reach beyond the maximum value of $\beta$ have no effect on the posterior, so it is always better to have too many bins than too few.

Because the Poisson distribution only has one well-defined parameter, it is computationally fast to calculate the posterior. As such, we have decided to solve for the posterior using 11 different bin sizes between 0.044 and 0.058. Then we add up all of the resulting posteriors and divide by 11 to normalize it once again. The final result is a less noisy posterior. In the bottom panel of Figure \ref{fig_example}, we plot the 11 individual posteriors in gray and the averaged posterior in blue. We take the median of the posterior to be the binary fraction, because this was shown shown in \citetalias{spencer2017b} to best reproduce the binary fraction in mock galaxies.

A final note worth mentioning is that while this method is similar to that in \citetalias{spencer2017b}, there are two key changes. The first is that we use a much smaller bin size, and continue binning up to the largest value of $\beta$. The previous method only used a total of six bins and lumped all $\beta$'s larger than 4 into one bin. One consequence of this was that we needed to fit a skewed normal function to $N(x|j)_{mod}$ rather than the much simpler Poisson that we have used here. Second, our earlier method in \citetalias{spencer2017b} used only one bin size to find the posterior, whereas we have taken the average of 11 here. Without this addition, the posteriors from \citetalias{spencer2017b} could shift a few percent to the right or left as a result of the wide binning. As we will see in Section \ref{sec_results}, both methods produce the same binary fraction for Leo II. However, the slight variability of binary fraction with bin size seen in our earlier analysis \citepalias{spencer2017b} becomes negligible using our new methodology.

\subsection{Repeatability}\label{sec_posterior}

\begin{figure}
\epsscale{1}
\plotone{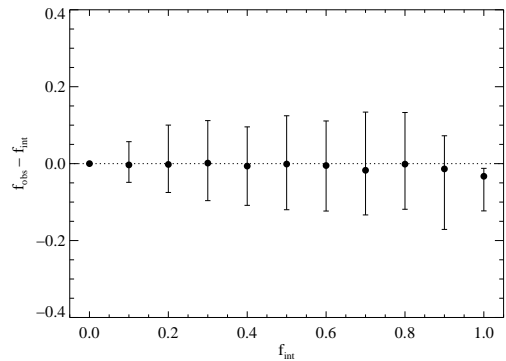}
\caption{The median observed binary fraction, $f_{obs}$, for 500 mock Draco-like galaxies is plotted against the intrinsic binary fraction, $f_{int}$. The horizontal dotted line is where the observed binary fraction perfectly matches the intrinsic binary fraction. Error bars show the range of observed binary fractions that 68\% of all mock galaxies fell between.}
\label{fig_mockgals}
\end{figure}

\begin{figure*}[t]
\epsscale{0.9}
\plottwo{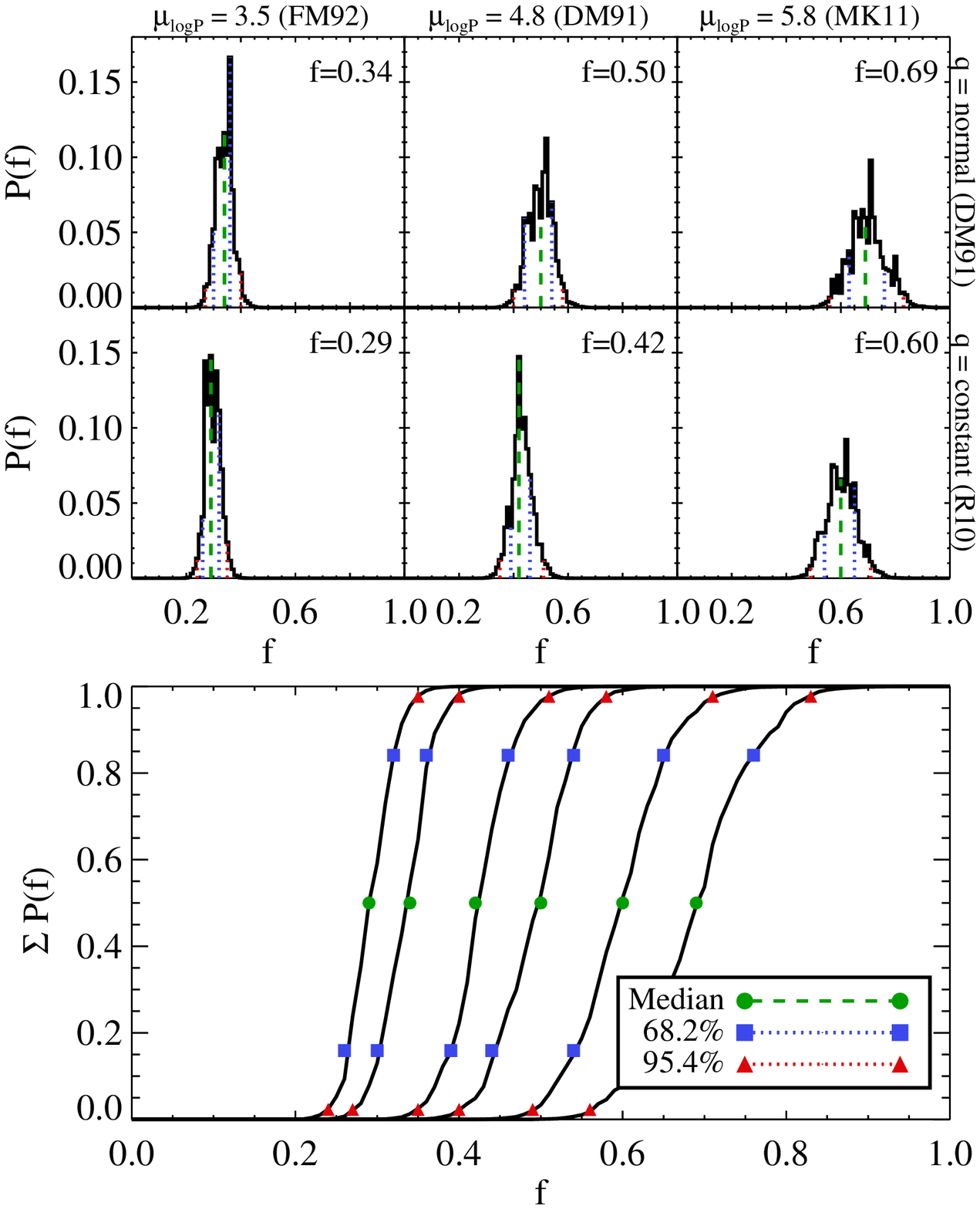}{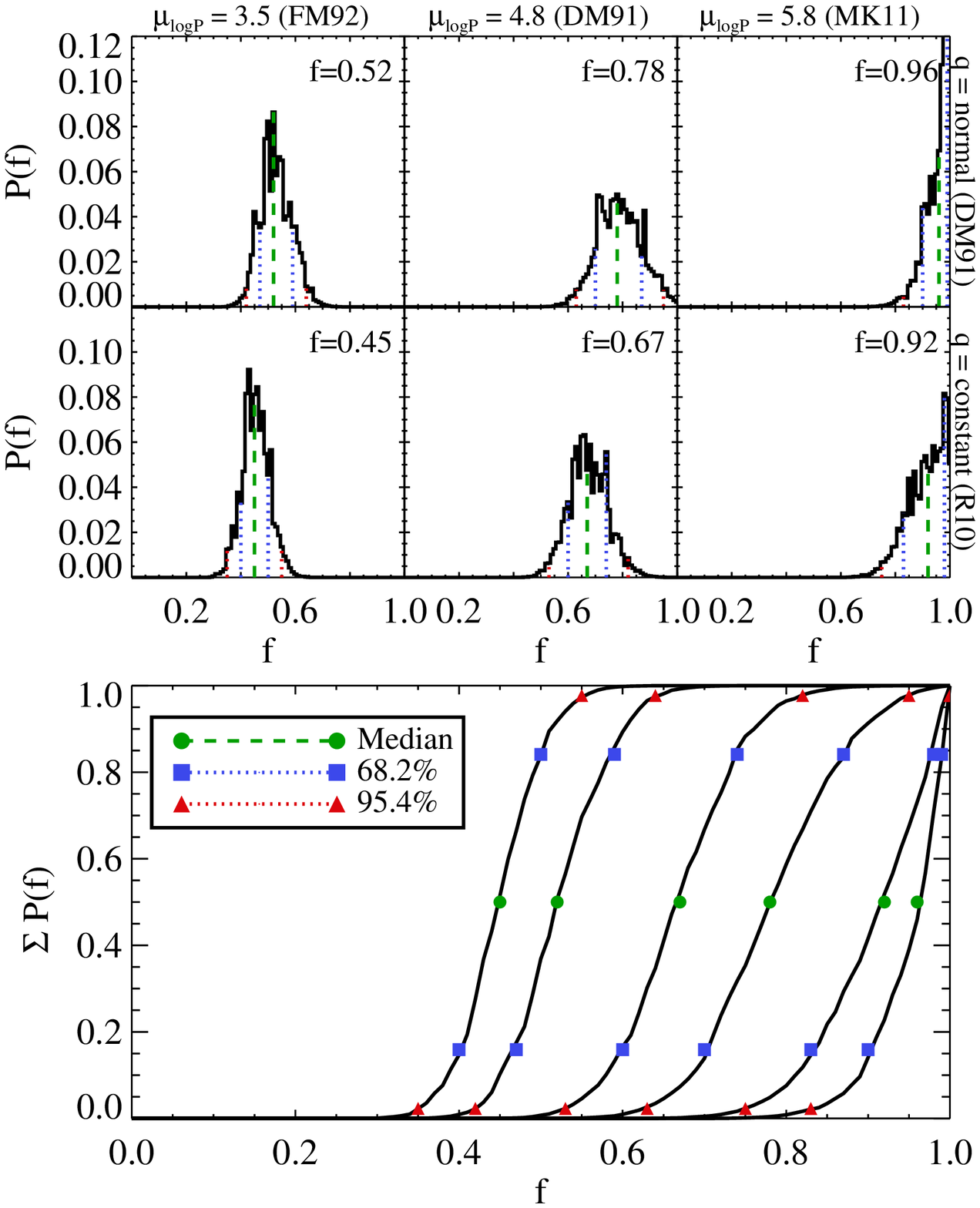}
\caption{Top panels are PPDs and bottom panels are cumulative PPDs for Draco (left) and Ursa Minor (right). The parameters used in the simulations are listed on the top and right axes of the PPDs. DM91=\citet{duquennoy1991}, FM92 = \citet{fischer1992}, R10 = \citet{raghavan2010}, MK11 = \citet{marks2011}.}
\label{fig_ppds}
\end{figure*}

To ensure that our method produces accurate results, we applied it to a series of test cases where the binary fraction was known. We considered 500 MC realizations for 11 binary fractions that were evenly spaced from $f=0$ to $f=1$. These mock galaxies were based on the velocity errors and observation times from the Draco data set. In all cases, we adopted a Gaussian mass ratio distribution \citepalias{duquennoy1991} and a log-normal period distribution centered on $\mu_{\log P}=4.8$ \citepalias{duquennoy1991}.  In Figure \ref{fig_mockgals}, we plot the difference between the observed binary fraction that our method recovered and the intrinsic binary fraction that was programmed into the galaxy. Black dots indicate the median observed binary fraction from the 500 mock galaxies. The error bars indicate the range that includes 68\% of the galaxies. The horizontal dotted line is the expectation, and it is indeed what we find for the majority of the mock galaxies. The only exception is the case where $f=1$, and presumably other very high binary fractions ($f\gtrsim0.9$), where the binary fraction is consistently underestimated by a few percent. Since it is physically unlikely for the binary fraction of an old stellar population to be near 1 \citep[][and references therein]{goodwin2006}, this discrepancy at high values is not a significant problem for realistic cases.

\section{Results}\label{sec_results}

Six combinations of mass ratio and period distributions were used to generate six complete sets of MC simulations, and consequently six posterior probability distributions (PPDs) of the binary fraction for each dSph. 
In Figure \ref{fig_ppds}, we plot all of the PPDs for Draco and Ursa Minor, respectively. The top row shows the posteriors with a normal mass ratio distribution \citepalias{duquennoy1991}, and the middle row has a constant mass ratio distribution \citepalias{raghavan2010}. The left column uses a log-normal period centered at $\mu_{\log P}=3.5$ \citepalias{fischer1992}, the middle column is centered at $\mu_{\log P}=4.8$ \citepalias{duquennoy1991}, and right column is centered at $\mu_{\log P}=5.8$ \citepalias{marks2011}. The median of each posterior, which we adopt as the binary fraction, is listed in the top right corner of each panel. 

As expected, the binary fraction is larger for higher values of $\mu_{\log P}$. Changes to the period distribution also have the largest effect on the posterior. Increasing $\mu_{\log P}$ from 3.5 to 5.8 increases the binary fraction by about $30\%$ for Draco and about $40\%$ for Ursa Minor. Alternatively, the mass ratio distributions we sampled from can only change the resulting binary fraction by 5-10\% for a given period distribution.

\begin{deluxetable*}{l c c c c c c c}
\centering
\tablewidth{0pt}
\tablecaption{Median and credible intervals of PPDs for Draco and Ursa Minor\label{ch4_table_ppds}}
\tabletypesize{\scriptsize}
\tablehead{\colhead{Galaxy} & \colhead{$q$ Distribution} & \colhead{$\sigma_{\log P}$ (log days)} &  \colhead{$\mu_{\log P}$ (log days)} & \colhead{$e$ Distribution} & \colhead{Median ($f$)} & \colhead{68.2\% Interval} & \colhead{95.4\% Interval}}
\startdata
     Draco &   normal (DM91) & 2.3 (DM91) & 4.8 (DM91) &   constant (R10) & 0.50 & 0.44-0.54 & 0.40-0.58 \\
     Draco &   normal (DM91) & 2.3 (DM91) & 3.5 (FM92) &   constant (R10) & 0.34 & 0.30-0.36 & 0.27-0.40 \\
     Draco &   normal (DM91) & 2.3 (DM91) & 5.8 (MK11) &   constant (R10) & 0.69 & 0.63-0.76 & 0.56-0.83 \\
     Draco &  constant (R10) & 2.3 (DM91) & 4.8 (DM91) &   constant (R10) & 0.42 & 0.39-0.46 & 0.35-0.51 \\
     Draco &  constant (R10) & 2.3 (DM91) & 3.5 (FM92) &   constant (R10) & 0.29 & 0.26-0.32 & 0.24-0.35 \\
     Draco &  constant (R10) & 2.3 (DM91) & 5.8 (MK11) &   constant (R10) & 0.60 & 0.54-0.65 & 0.49-0.71 \\
\hline
Ursa Minor &   normal (DM91) & 2.3 (DM91) & 4.8 (DM91) &   constant (R10) & 0.78 & 0.70-0.87 & 0.63-0.95 \\
Ursa Minor &   normal (DM91) & 2.3 (DM91) & 3.5 (FM92) &   constant (R10) & 0.52 & 0.47-0.59 & 0.42-0.64 \\
Ursa Minor &   normal (DM91) & 2.3 (DM91) & 5.8 (MK11) &   constant (R10) & 0.96 & 0.90-0.99 & 0.83-1.00 \\
Ursa Minor &  constant (R10) & 2.3 (DM91) & 4.8 (DM91) &   constant (R10) & 0.67 & 0.60-0.74 & 0.53-0.82 \\
Ursa Minor &  constant (R10) & 2.3 (DM91) & 3.5 (FM92) &   constant (R10) & 0.45 & 0.40-0.50 & 0.35-0.55 \\
Ursa Minor &  constant (R10) & 2.3 (DM91) & 5.8 (MK11) &   constant (R10) & 0.92 & 0.83-0.98 & 0.75-1.00 \\
\enddata
\tablecomments{DM91 = \citet{duquennoy1991}, FM92 = \citet{fischer1992}, R10 = \citet{raghavan2010}, MK11 = \citet{marks2011}.}
\end{deluxetable*}

Certain clear correlations arise for specific adopted parameters. For example, the smallest $f$ in both Draco and Ursa Minor is found with a constant mass ratio distribution \citepalias{raghavan2010} and a log-normal period distribution with a location of $\mu_{\log P}=3.5$ \citepalias{fischer1992}. The largest binary fraction for both galaxies corresponds to a normal mass ratio distribution \citepalias{duquennoy1991} and a log-normal period distribution with $\mu_{\log P}=5.8$ \citepalias{marks2011}. The smallest and largest binary fractions found in Draco are $0.29^{+0.03}_{-0.03}$ and $0.69^{+0.07}_{-0.06}$. For Ursa Minor, the binary fraction ranges from $0.45^{+0.05}_{-0.05}$ to $0.96^{+0.03}_{-0.06}$. Although the binary fractions vary considerably with the binary orbital parameters, we can still rule out $f>0.86$ and $f<0.22$ in Draco with 99\% confidence. Similarly, binary fractions below $0.32$ can be ruled out with 99\% confidence in Ursa Minor. It should be noted that while these limits do depend on binary orbital parameters, it is not likely that the binary fraction will be beyond these limits, because we specifically chose parameter distributions that explored the largest range of observed parameters. A full summary of the PPDs is provided in Table \ref{ch4_table_ppds}. 

\begin{figure}
\plotone{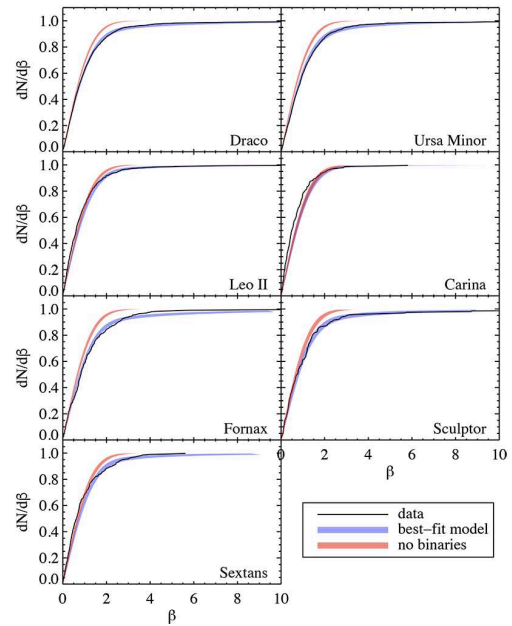}
\caption{Cumulative histograms of $\beta$ for each of the seven dSphs. The observations are a black line. The blue envelope encloses 68\% of the Monte Carlo realizations for the best fitting binary fraction. The red envelope encloses 68\% of the Monte Carlo realizations for zero binaries.}
\label{fig_ref}
\end{figure}

Note that the distribution of $\beta_{obs}$ from the observations do not perfectly match the distributions of $\beta_{mod}$ from the simulations. This can be seen in the cumulative distributions of $\beta$ in Figure \ref{fig_ref}. $\beta_{obs}$ is shown as a black line, the envelope enclosing 68\% of the $\beta_{mod}$ from the MC models with the best fitting binary fraction is shown in blue, and the red envelope encloses 68\% of the $\beta_{mod}$ for the case of zero binaries. The top two panels are for Draco and Ursa Minor, and the remaining panels are for additional dSphs that we consider in the next section. The discrepancy between the observations and the best-fit models are more pronounced in some of the other galaxies than they are for Draco and Ursa Minor. Taken as a whole, this suggests that the binary parameter combinations that we used do not reflect the actual parameters found in the dwarfs. Although beyond the scope of this work, it seems possible that some constraints could be put on the mass ratio and period distributions by considering the shape of the $\beta$ distributions. The best we can do here is comment on which of the six parameter sets provides the best fit to the observations for the cases of Draco and Ursa Minor.

\begin{figure*}
\centering
\plottwo{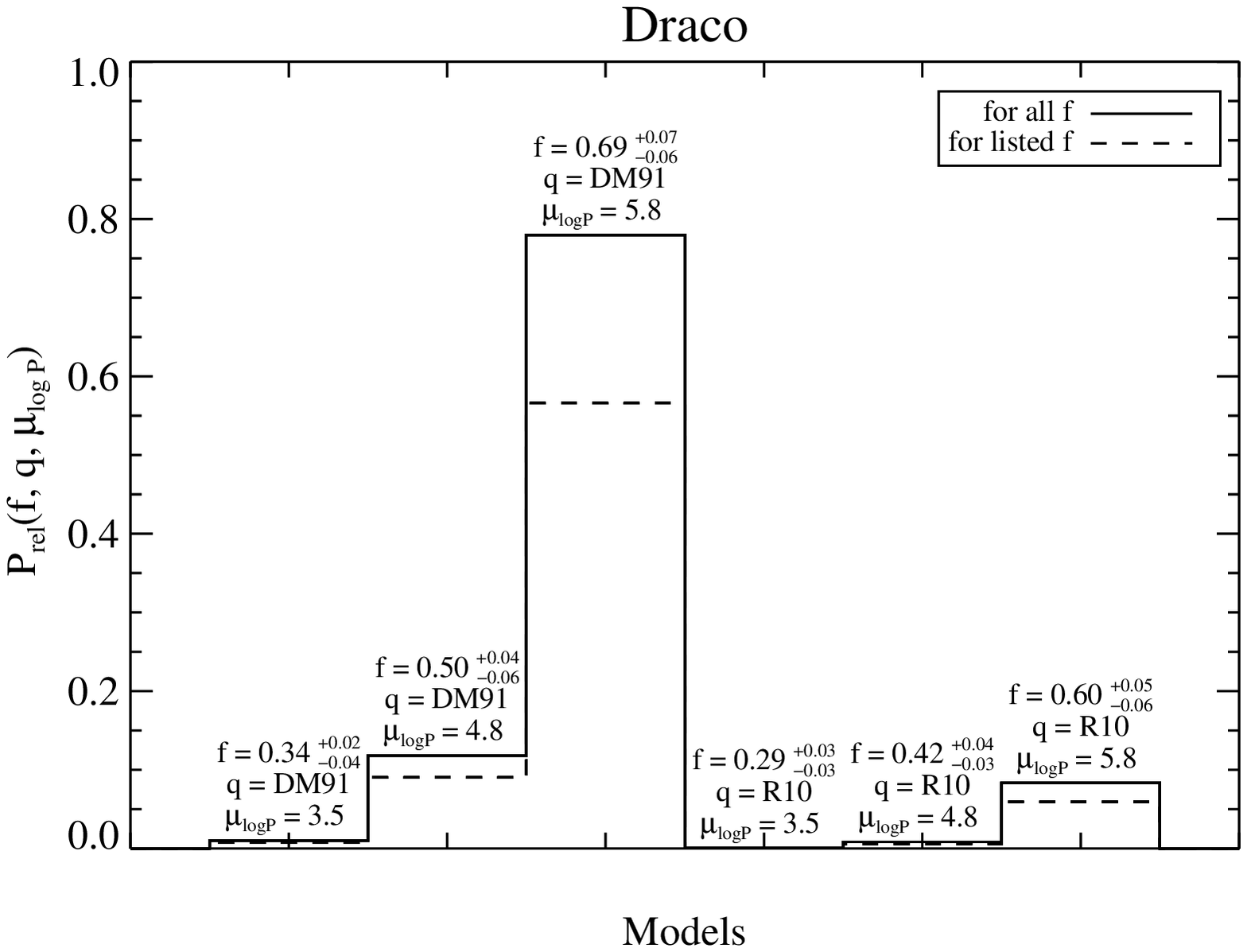}{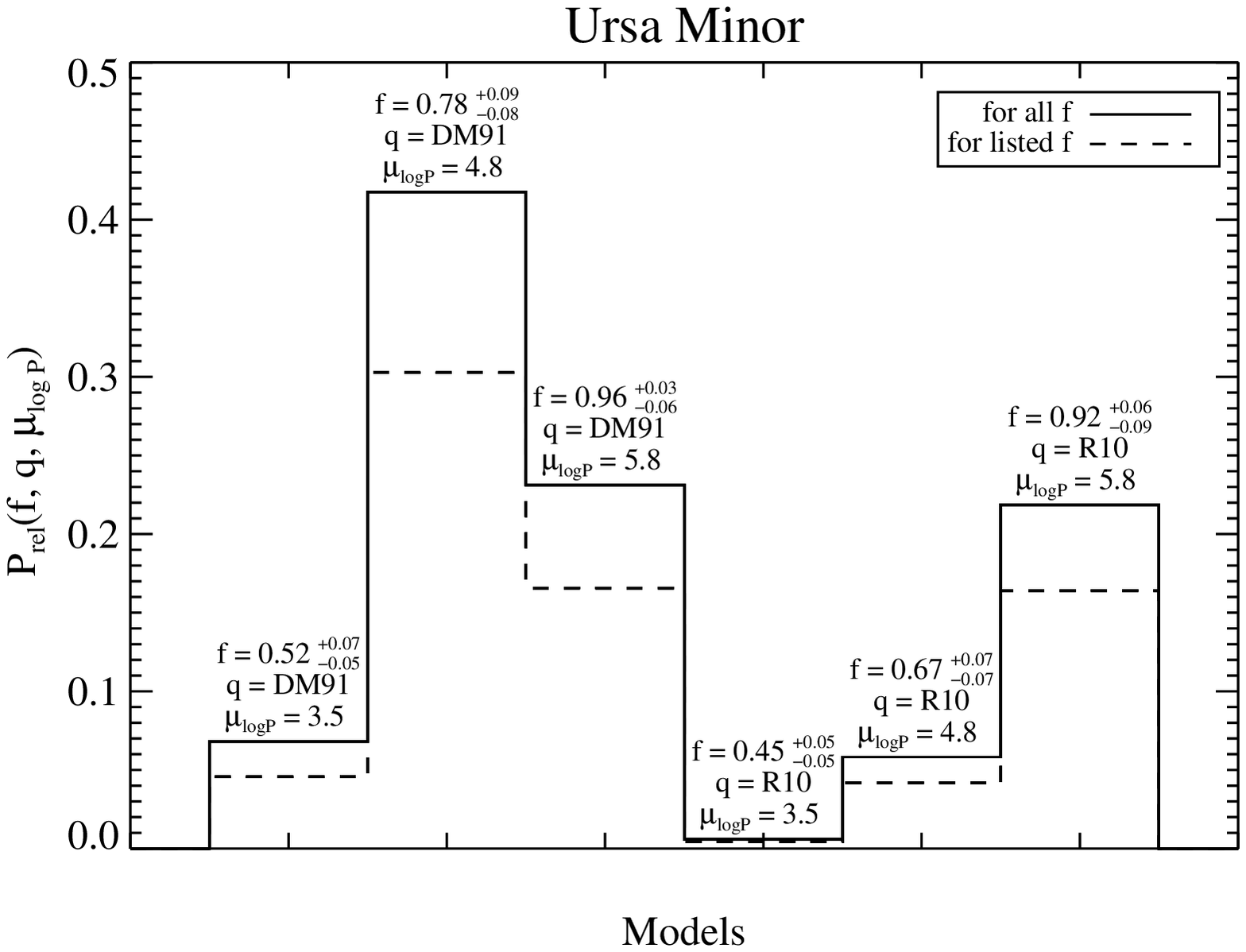}
\caption{Relative probability of a model fitting the data. Model parameters are indicated above each bar. Bars are normalized so that the sum is equal to 1. Solid lines compare the relative probabilities of the PPDs over all values of $f$ ($0\le f\le1$), and dashed lines are for the values of $f$ within the 68\% credible intervals (listed above each bar). The left panel is Draco, and the right panel is Ursa Minor.}
\label{fig_which_params}
\end{figure*}

We repeat the Bayesian analysis from Section \ref{sec_mcsims} that was used to generate the PPDs in Figure \ref{fig_ppds}, with one adjustment. We normalize the posterior such that the sum of all six PPDs for each of the models is equal to one, rather than the sum of each individual PPD being equal to one. Because these are relative probabilities, their values have no physical meaning, but comparisons between models can be used to say which model is more likely, and by how much. In Figure \ref{fig_which_params}, the relative probability for each of the six models over all values of $f$ is shown as a solid line; the relative probability over the 68\% credible interval is shown as a dashed line. Parameter distributions are listed above each bar in the figure.

In all cases, the models with the \citetalias{duquennoy1991} mass ratio distribution were more likely than the models with the \citetalias{raghavan2010} mass ratio distribution for any given period distribution. For a given mass ratio distribution, the models with $\mu_{\log P}=3.5$ (as in \citetalias{fischer1992}) always had the lowest probability. The set of parameters that best reproduced the data in Draco was the \citetalias{duquennoy1991} mass ratio distribution and a period distribution with $\mu_{\log P}=5.8$, which corresponded to a binary fraction of $0.69^{+0.07}_{-0.06}$. For Ursa Minor, the best parameters were the \citetalias{duquennoy1991} mass ratio distribution and \citetalias{duquennoy1991} period distribution ($\mu_{\log P}=4.8$), which had a binary fraction of $0.78^{+0.09}_{-0.08}$. The three best fitting models---\citetalias{duquennoy1991} mass ratio distribution with either $\mu_{\log P}=4.8$ or $\mu_{\log P}=5.8$, and \citetalias{raghavan2010} mass ratio distribution with $\mu_{\log P}=5.8$---are the same for both dwarfs. While we cannot comment on the absolute parameters of the mass ratio and period distributions, we can say that the \citetalias{duquennoy1991} mass ratio distribution is preferred over the \citetalias{raghavan2010} distribution, and that the period distribution peaks toward longer periods. A more continuous exploration of the parameter distributions should yield better constraints on the period and mass ratio distributions. Since the \citetalias{duquennoy1991} parameter distributions are preferred by Ursa Minor and are more commonly found in other binary literature works, we use those parameters in the discussion that follows.

\subsection{Binary Fractions among Dwarfs}\label{sec_dwarfs}

Binary fractions for Carina, Fornax, Sculptor, and Sextans have previously been reported by \citet{minor2013} based on MMFS/Magellan data from \citet{walker2009a}. These data spanned about 1 year and had 2--4 repeat observations. Due to the limits on $n$ and $\Delta t_{max}$, that data set was not ideal for a binary fraction analysis, but it still proved suitable to produce broad PPDs that ruled out some binary fractions. Given these results, it was natural to consider combining them with our results for Draco, Ursa Minor, and Leo II \citep{spencer2017b} to explore the behavior of the binary fraction across the more luminous MW dSph systems. However, there are some differences between our methods and those in \citet{minor2013} that make a simple combination of results problematic and potentially misleading. For example, they used a different eccentricity distribution, applied an error model to their analysis, and performed the Bayesian analysis on a star-by-star basis rather than as a data set, all in contrast to the approach we describe here. Since a comparison still seems desirable, we chose to apply our methodology to the MMFS/Magellan data used by \citet{minor2013} to ensure consistency among binary fraction calculations.

\begin{deluxetable}{l c c c c}
\centering
\tablewidth{0pt}
\tablecaption{Quantities used to derive $a_{\mathrm{max}}$ in seven dSphs\label{table_amax}}
\tabletypesize{\scriptsize}
\tablehead{\colhead{Galaxy} & \colhead{$v_{sys}$\tablenotemark{a}} & \colhead{$\sigma$\tablenotemark{b}} & \colhead{Source of Data\tablenotemark{c}} & \colhead{$I_0$\tablenotemark{d}}\\
 & \colhead{(\kms{})} & \colhead{(\kms{})} & & \colhead{($L_{\odot}~\mathrm{pc}^{-3}$)}}
\startdata
Draco & $-292.3\pm0.4$ & $9.0\pm0.3$ & \citet{walker2015} & 0.008 \\
Ursa Minor & $-246.9\pm0.4$ & $8.0\pm0.3$ & Tab\ref{tab:umi_table1} & 0.006 \\
Leo II & $78.5\pm0.6$ & $7.4\pm0.4$ & \citet{spencer2017a} & 0.029 \\
Carina & $223.0\pm0.3$ & $6.4\pm0.3$ & \citet{walker2009a} & 0.006 \\
Fornax & $54.9\pm0.2$ & $11.8\pm0.2$ & \citet{walker2009a} & 0.018 \\
Sculptor & $111.3\pm0.2$ & $8.4\pm0.1$ & \citet{walker2009a} & 0.055 \\
Sextans & $224.0\pm0.4$ & $8.2\pm0.4$ & \citet{walker2009a} & 0.002 \\
\enddata
\tablenotetext{a}{Systemic velocity}
\tablenotetext{b}{Velocity dispersion}
\tablenotetext{c}{Source for velocity data that we used to determine the systemic velocity and velocity dispersion.}
\tablenotetext{d}{Central luminosity density from \citet{mateo1998}.}
\end{deluxetable}

\begin{deluxetable}{l c c c}
\centering
\tablewidth{1000pt}
\tablecaption{Binary fractions for seven dSphs\label{table_ppddwarfs}}
\tabletypesize{\scriptsize}
\tablehead{\colhead{Galaxy} & \colhead{$f$} & \colhead{$f_{\mathrm{ref}}$} & \colhead{Reference}}
\startdata
Draco & $0.50^{+0.05}_{-0.04}$ &  -  & - \\
Ursa Minor & $0.78^{+0.08}_{-0.09}$ &  -  & - \\
Leo II & $0.36^{+0.07}_{-0.08}$ & $0.33^{+0.12}_{-0.09}$ & \citet{spencer2017b} \\
Carina & $0.20^{+0.09}_{-0.13}$ & $0.14^{+0.28}_{-0.05}$ & \citet{minor2013} \\
Fornax & $0.87^{+0.12}_{-0.09}$ & $0.44^{+0.26}_{-0.12}$ & \citet{minor2013} \\
Sculptor & $0.58^{+0.15}_{-0.17}$ & $0.59^{+0.24}_{-0.16}$ & \citet{minor2013} \\
Sextans & $0.71^{+0.15}_{-0.14}$ & $0.69^{+0.19}_{-0.23}$ & \citet{minor2013} \\
\enddata
\end{deluxetable}

\begin{figure}
\epsscale{1.1}
\plotone{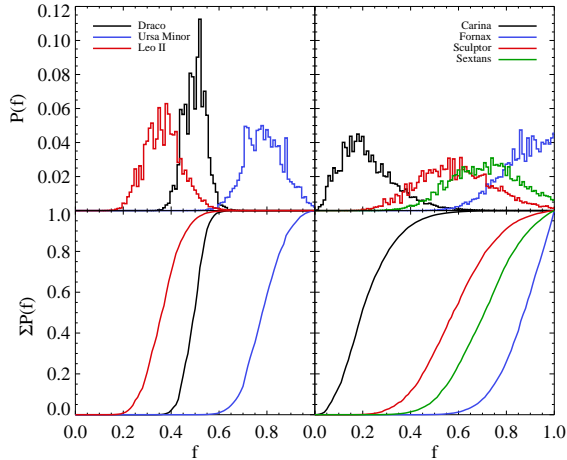}
\caption{Top: PPDs for seven dwarfs using a normal mass ratio distribution and a log-normal period distribution centered on $\mu_{\log P}=4.8$. Bottom: cumulative distributions of the posterior.}
\label{fig_ppdall}
\end{figure}

The MMFS/Magellan data set contains both member and nonmember stars of their respective dwarf galaxies \citep{walker2009a}. We chose to select stars as members if they were within three times the velocity dispersion of the systemic velocity. The velocity dispersions and systemic velocities that we derived using the method from Section \ref{sec_membership} are listed in Table \ref{table_amax}. These velocity dispersions were also used in conjunction with the luminosity densities from \citet{mateo1998} to derive $a_{max}$. We assume that the reported velocity error measurements are accurate. (Systematically overestimated errors would lead to an underestimated binary fraction, whereas underestimated errors would lead to an overestimated binary fraction.) The results of our simulations for a normal mass ratio distribution and log-normal period distribution with $\mu_{\log P}=4.8$ (i.e., parameters equivalent to those in the top-center panels of Figure \ref{fig_ppds}) agree very well with those from \citet{minor2013} in three of the four cases. Our binary fractions for Carina, Fornax, Sculptor, and Sextans are $0.20^{+0.09}_{-0.13}$, $0.87^{+0.12}_{-0.09}$, $0.58^{+0.15}_{-0.17}$, and $0.71^{+0.15}_{-0.14}$, respectively. Those from \citet{minor2013} are $0.14^{+0.28}_{-0.05}$, $0.44^{+0.26}_{-0.12}$, $0.59^{+0.24}_{-0.16}$, and $0.69^{+0.19}_{-0.23}$. These binary fraction results are displayed in Table \ref{table_ppddwarfs}.

The discrepancy between values for Fornax is almost certainly due to the treatment of the velocity errors. \citet{minor2013} estimated that the velocity errors on Fornax were under-reported by a factor of 55\%. We can also see that the histogram of $P(\chi^2,\kappa)$ for Fornax exhibits some strange behavior. The number of stars per bin is not uniformly biased toward low $P$, as one would expect for a systematic underestimate of the velocity errors. Rather, there are some bins in the middle of the distribution that contain more values than expected by Poisson errors. Since we are not set up to treat improperly reported velocity errors in our simulations, it is not surprising that our results are very different from what was previously reported. We ran two additional simulations in which we applied a constant corrective factor to the velocity errors. In the first case we multiplied the errors by a factor of 1.55 to match the estimates from \citet{minor2013}, which yielded a binary fraction of $0.22^{+0.11}_{-0.09}$. In the second case we multiplied by a factor of 1.15 because we found this provided the best correction to our $P(\chi^2)$ histogram. It yielded a binary fraction of $0.61^{+0.15}_{-0.13}$.

It is curious that \citet{minor2013} finds such a large discrepancy between the reported velocity errors \citep{walker2009a} and their own velocity error estimates for Fornax but not for the other dwarfs, because all of the data were taken on the same instrument and often during the same run. An alternative to the Fornax errors being largely underestimated is that Fornax actually has a large binary fraction. Binary fraction and velocity errors are somewhat hard to disentangle. If the errors have been underestimated, then the binary fraction will appear large; if the binary fraction is large, then the errors will appear to be underestimated. The best way to determine velocity errors in the context of binaries is on a nightly basis by comparing the measurements from multiple exposures. Or if that is not possible, then using exposures taken over the course of a couple nights should suffice. Velocities observed over such short timescales should not have any significant velocity variability caused by binaries and should represent the observational errors.

A binary fraction for Leo II was reported in \citetalias{spencer2017b}, but since the method in that chapter was slightly different than the one here, we chose to run a new set of MC simulations for Leo II as well. We used the normal mass ratio distribution and log-normal period distribution with $\mu_{\log P}=4.8$, as was done for Carina, Fornax, Sculptor, and Sextans. The binary fraction for Leo II came out to be $0.36^{+0.07}_{-0.08}$, in good agreement with the previous results. The posteriors for all seven dwarfs are plotted in Figure \ref{fig_ppdall}, and the binary fractions are listed in Table \ref{table_ppddwarfs}. 

The posteriors cover a large range of binary fractions, yet their distributions are wide and all overlap around 0.55--0.60. This overlap region is small, suggesting that binary fraction is not a constant property across all dwarfs. Nevertheless, it is still valuable to determine the probability that the binary fraction is the same and, if it is the same, what value it takes on.

For the purpose of this discussion, we define ``the same'' as all the binary fractions being within some specified range. The width, $w$, of that range can be any value, but we chose to focus on $w=0.1$ and $w=0.2$. These were selected because the 68\% credible interval for Draco was $\approx0.1$, and for Ursa Minor, Leo II, Carina, and Fornax, it was $\approx0.2$. 

We can calculate the 7-dimensional joint probability that all the dwarfs have a binary fraction within some width, $w$, centered on some binary fraction $f_g$. (For example, the probability that all the dwarfs have a binary fraction between 0.4 to 0.6 would be the case where $f_g=0.5$ and $w=0.2$.) First, we take the sum of the PPD for a single dSph over the range $(f_g-w/2)$ to $(f_g+w/2)$. This can be written as 
\begin{equation*}
P_d(w,f_g)=\sum_{f=f_g-w/2}^{f_g+w/2}P_d(f), 
\end{equation*}
which yields the probability that the binary fraction for the dwarf, $d$, is within the specified range, $w$,  centered on some binary fraction, $f_g$. Since we have normalized $P_d(f)$ such that the sum over all $f$ is equal to 1.0, this term will always be $<1$. 

\begin{figure}
\epsscale{1}
\plotone{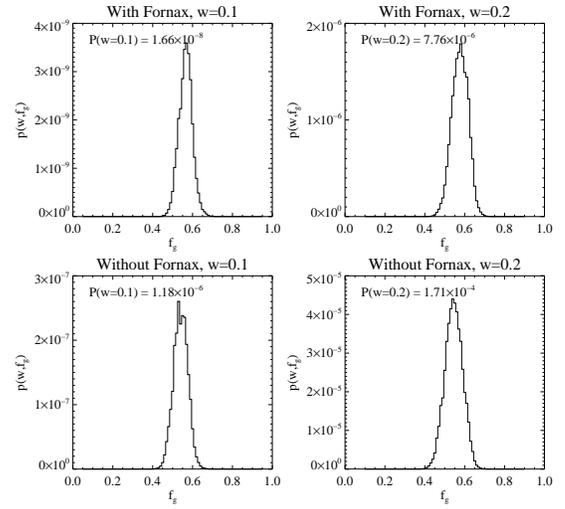}
\caption{Probability that the binary fraction for all galaxies exists within a finite range, $w$, that is centered on $f_g$, as described by Equation \ref{eq_pwf}. The left two panels use $w=0.1$, and the right two use $w=0.2$. The top two panels include seven galaxies, and the bottom two include six galaxies. The total probability that binary fraction exists within some width $w$ regardless of $f_g$ is printed in the top left corner of each panel, and is defined in Equation \ref{eq_pw}. This figure was made under the assumptions that the period and mass ratio distributions take the forms described in \citet{duquennoy1991}.}
\label{fig_onebf}
\end{figure}

To find the probability that the binary fraction is within the range $w$ centered on $f_g$ for all the dwarfs, we need only take the product of the sums over over $d$. Assuming the PPDs are independent, this is given by
\begin{equation}
P(w,f_g) = \prod_{d}P_d(w,f_g)= \prod_{d}\Bigg(\sum_{f=f_g-w/2}^{f_g+w/2}P_d(f)\Bigg),
\label{eq_pwf}
\end{equation}
where $d$ is the set of dwarfs, $d=\{$Draco, Ursa Minor, Leo II, Carina, Fornax, Sculptor, Sextans$\}$, and $P_d(f)$ is the PPD corresponding to that dwarf. We plot this probability as a function of $f_g$ in Figure \ref{fig_onebf}, with $w=0.1$ occupying the left two panels and $w=0.2$ in the right two panels. Since our posterior for Fornax did not agree with the previously published value \citep{minor2013}, we do this for the sample of seven dwarfs (top two panels), and for a sub-sample that excludes Fornax (the bottom two panels). Due to the formulation of Equation \ref{eq_pwf}, there are only values of $P(w,f_g)$ in the range of $0.05\leq f_g \leq0.95$ for $w=0.1$, and in the range of $0.1\leq f_g \leq0.9$ for $w=0.2$. (For example, if we selected $f_g=0$ and $w=0.1$, then the lower limit on the sum in Equation \ref{eq_pwf} would be $f=-0.05$, which is not a physically possible value for the binary fraction.)

The foremost feature of Figure \ref{fig_onebf} is the extremely small probabilities along the \textit{y}-axes, which range from $10^{-9}$ to $10^{-5}$. These values imply that it is unlikely for the binary fraction to be ``the same'' (i.e., within a 20\% range) for the dSphs considered here. This concept will be given additional attention later in this section when we introduce another form of the probability equation.

The maximum probability for the seven-galaxy sample occurs at $f_g=0.57$ for $w=0.1$ and $f_g=0.58$ for $w=0.2$. This means that if these dSphs have binary fractions within 0.1 (0.2) of each other, this is most likely to occur in the range $0.52\leq f\leq0.62$ ($0.48\leq f\leq0.68$). For the six-galaxy sample, the maximum probability occurs at $f_g=0.53$ ($f_g=0.54$) for $w=0.1$ ($w=0.2$). This means that if these dSphs have binary fractions within a range of 0.1 (0.2), then it is most likely to occur when $0.48\leq f\leq0.58$ ($0.44\leq f\leq0.64$).

The range of binary fractions for the sample of seven galaxies spans higher values than the sample of six galaxies because our analysis finds a large binary fraction for Fornax. As a result, Fornax imposes a lower limit on $f_g$. When Fornax is removed, then $f_g$ can shift toward lower values, but is still limited by Ursa Minor. On the other end, Carina and Leo II impose an upper limit on $f_g$. 

The sample excluding Fornax has higher---though still very small---probabilities of the binary fraction being the same, as can be seen by the \textit{y}-axis labels in Figure \ref{fig_onebf}. This is once again because our analysis finds a large binary fraction for Fornax. The probability of Fornax having $f<0.6$ is only 1\%, and when such small numbers get multiplied through Equation \ref{eq_pwf}, the result is very small probabilities. These probabilities are about two orders of magnitude smaller than the probabilities that exclude Fornax. To summarize, the inclusion of Fornax (1) pulls $f_g$ toward higher values, and (2) decreases the probability that $f$ could be the same for all dwarfs. Regardless of whether or not Fornax is included, the probability that the binary fraction for all the dwarfs is ``the same'' in these intervals is extremely small.

We now turn to a new question: how large must $w$ be for the probability of $f$ being ``the same'' to become appreciable? We tackle this by summing $P(w,f_g)$ over all $f_g$ and exploring a continuous choice of $w$. The probability that all the dwarfs have a binary fraction within some range, $w$, centered on any value of $f_g$ can be expressed as

\begin{equation}
\begin{split}
P(w) = & \prod_d\bigg(\sum_{f=0}^{w}P_d(f)\bigg) + \\ 
& \sum_{{f_g=0.01+w/2}}^{1-w/2}\Bigg(\prod_d\bigg(\sum_{f=f_g-w/2}^{f_g+w/2}P_d(f)\bigg)- \\
& \prod_d\bigg(\sum_{f=f_g-w/2}^{f_g+w/2-0.01}P_d(f)\bigg)\Bigg)
\end{split}
\label{eq_pw}
\end{equation}
The last product term is subtracted to prevent some probabilities from being counted more than once.

\begin{figure}[t]
\epsscale{1}
\plotone{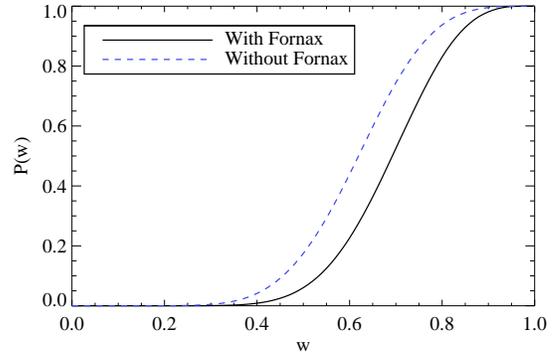}
\caption{Probability that the binary fractions of dSphs exist within a specified range with width $w$. The solid black line includes seven galaxies; the dashed blue line includes six galaxies. The dwarfs do not occupy the same range of $f$ until that range is widened to about 0.3 or 0.4. This figure was made under the assumptions that the period and mass ratio distributions take the forms described in \citet{duquennoy1991}. }
\label{fig_onebf_p95}
\end{figure}

Figure \ref{fig_onebf_p95} plots this probability as a function of $w$. The set of seven galaxies is shown by the black solid line, and the set of six galaxies is shown by the blue dashed line. The probability of the six-galaxy sample becomes greater than 1\% around $w=0.3$. For the seven-galaxy sample, this transition occurs around $w=0.4$. This means that the binary fractions of the galaxies do not all begin to occur within some specified range until that range has a width of at least 0.3--0.4 in $f$. This is larger than the credible intervals of most of the PPDs (as in Table \ref{table_ppddwarfs}), so an alternative interpretation is that the binary fractions should no longer be considered ``the same'' when $w$ is this large. Rather, the binary fractions are spread over some range of values with a width of at least 0.3--0.4. 

We produced variations of Figure \ref{fig_onebf_p95} for different period and mass ratio distributions and found that the \citet{fischer1992} distribution could bring this turning point down to as low as $w=0.2$. Regardless of the inclusion of Fornax or binary parameter distributions, there is a $<1\%$ chance that the binary fractions for the considered dSphs all exist within some range of $f$ with width 0.2. Ultimately, we find that it is highly unlikely that the binary fraction is constant across dwarf spheroidal galaxies. The only other way in which the binary fraction could be the same is if the period distributions varied among the dwarfs, with dwarfs like Leo II and Carina have longer average periods and dwarfs like Sextans, Ursa Minor, and Fornax having shorter mean periods. 

Assuming that binary fraction does vary among dwarfs and that the binary orbital parameters are constant, we examined whether binary fraction is dependent on any galactic properties. The properties we considered were distance from the Milky Way \citep{mcconnachie2012}, absolute magnitude \citep{mcconnachie2012}, surface brightness \citep{mateo1998}, luminosity density \citep{mateo1998}, mass density \citep{mateo1998}, total mass within half-light radius \citep{walker2009c}, velocity dispersion \citep{mcconnachie2012}, half-light radius \citep{mateo1998}, ellipticity \citep{mcconnachie2012}, mean metallicity \citep{kirby2011}, time to form 50\% of the stellar mass \citep{weisz2014}, and time to form 95\% of the stellar mass \citep{weisz2014}. Table \ref{table_dwarfprops1} lists the values for the properties for the eight classical dSphs. We compare these properties to the binary fractions of Draco, Ursa Minor, and Leo II that were calculated in this chapter, and to the binary fractions of Carina, Fornax, Sculptor, and Sextans in \citet{minor2013}.

\begin{figure*}
\epsscale{0.90}
\plotone{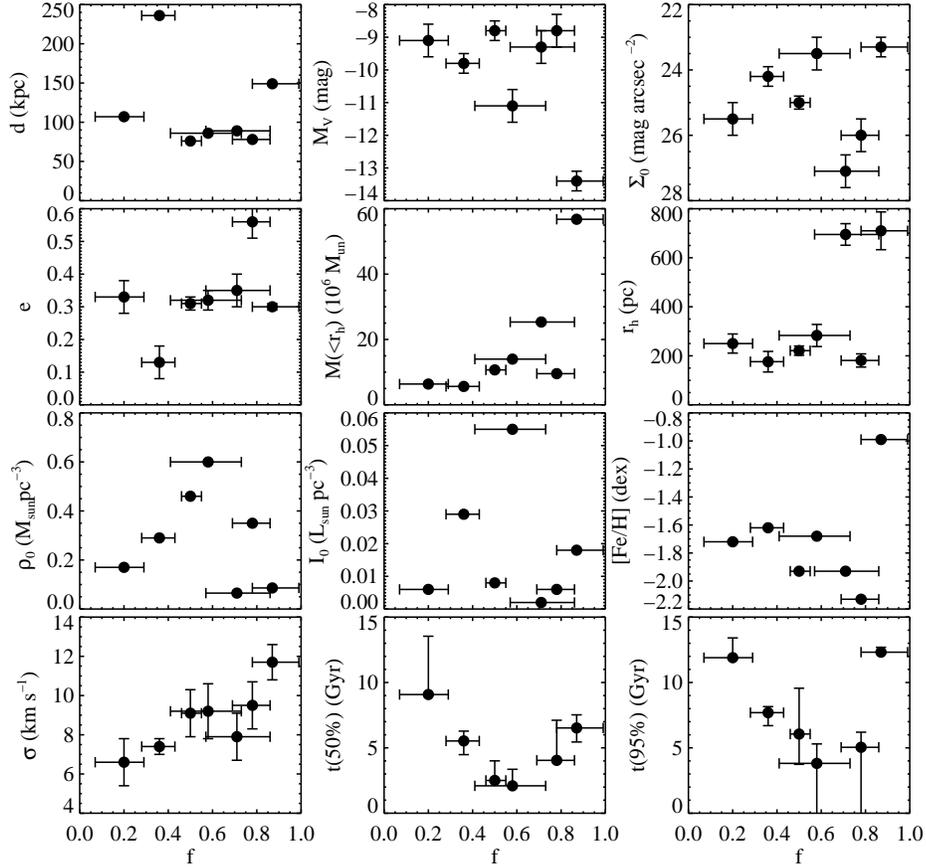}
\caption{Binary fraction, $f$, is plotted against other properties of the galaxies. Top row: distance, absolute V-band magnitude, V-band central surface brightness. Second row: ellipticity, total mass within half-light radius, half-light radius. Third row: central mass density, central luminosity density, mean metallicity. Bottom row: velocity dispersion, time to form 50\% of stars, time to form 95\% of stars.}
\label{fig_props}
\end{figure*}

Figure \ref{fig_props} shows all of these parameters plotted against binary fraction. Most cases yield scatterplots. The three parameters that exhibited the most promising correlations with binary fraction are velocity dispersion (bottom left panel), time since forming 50\% of the stellar mass (bottom middle panel), and time since forming 95\% of the stellar mass (bottom right panel). The loose trends that we find are that binary fraction roughly increases with velocity dispersion, and that galaxies that formed more of their stars early on have higher binary fractions than those with a more extended star formation history. The trend with velocity dispersion has increased significance if Fornax is included, while the trends with star formation history have decreased significance if Fornax is included.

Recall that one of the underlying purposes of this research is to see if binaries can alter our view of ultra-faints. The implication of the first trend is that ultra-faints would have low binary fractions. As a consequence, their velocity dispersions would have very minor inflation due to binaries. This seems unlikely given the cases of Bootes I \citep{koposov2011} and Segue 1 \citep{simon2011}, which did have 0.5-2 \kms{} velocity dispersion corrections due to binaries.

\begin{deluxetable*}{l c c c c c c c c c c c}[p]
\centering
\tablewidth{0pt}
\tablecaption{Properties of classical dSphs\label{table_dwarfprops1}}
\tabletypesize{\scriptsize}
\tablehead{\colhead{Dwarf}  & \colhead{Distance\tablenotemark{a}} & \colhead{M$_v$\tablenotemark{a}} & \colhead{Surf. bright\tablenotemark{a}} & \colhead{Ellipticity\tablenotemark{a}} & \colhead{r$_{\mathrm{half}}$\tablenotemark{a}} & \colhead{L$_v$\tablenotemark{b}} & \colhead{Lum. Density\tablenotemark{b}} & \colhead{$\sigma$\tablenotemark{a}} & \colhead{$M_{dyn}(\le r_{half})$\tablenotemark{a}} & \colhead{Mass Density\tablenotemark{b}}  & \colhead{Mean [Fe/H]\tablenotemark{a}} \\
 & \colhead{(kpc)} & \colhead{(mag)} & \colhead{(mag arcsec$^{-2}$)} & & \colhead{(pc)} & \colhead{$10^6 L_{\odot}$} & \colhead{($\mathrm{L}_{\odot} \mathrm{pc}^{-3}$)} & \colhead{km s$^{-1}$} & \colhead{$10^6 M_{\odot}$} & \colhead{($\mathrm{M}_{\odot}~\mathrm{pc}^{-3}$)} & \colhead{(dex)}}
\startdata
Draco	   	& 76	& -8.8$\pm$0.3	& 25.0$\pm$0.2 & 0.31$\pm$0.02 & 221$\pm$19 & 0.26 & 0.008  & 9.1$\pm$1.2  & 11  & 0.46  & -1.93$\pm$0.01 \\
Ursa Minor	& 78	& -8.8$\pm$0.5	& 26.0$\pm$0.5 & 0.56$\pm$0.05 & 181$\pm$27 & 0.29 & 0.006  & 9.5$\pm$1.2  & 9.5 & 0.35  & -2.13$\pm$0.01 \\
Leo II		& 236	& -9.8$\pm$0.3	& 24.2$\pm$0.3 & 0.13$\pm$0.05 & 176$\pm$42 & 0.58 & 0.029  & 7.4$\pm$0.4  & 4.6 & 0.29  & -1.62$\pm$0.01 \\
Leo I	  	& 258	& -12.0$\pm$0.3	& 22.6$\pm$0.3 & 0.21$\pm$0.03 & 251$\pm$27 & 4.79 & 0.092  & 9.2$\pm$1.4  & 12  & 0.28  & -1.43$\pm$0.01 \\
Carina		& 107	& -9.1$\pm$0.5	& 25.5$\pm$0.5 & 0.33$\pm$0.05 & 250$\pm$39 & 0.43 & 0.006  & 6.6$\pm$1.2  & 6.3 & 0.17  & -1.72$\pm$0.01 \\
Fornax	  	& 149	& -13.4$\pm$0.3	& 23.3$\pm$0.3 & 0.30$\pm$0.01 & 710$\pm$77 & 15.5 & 0.018  & 11.7$\pm$0.9 & 56  & 0.086 & -0.99$\pm$0.01 \\
Sculptor  	& 86	& -11.1$\pm$0.5	& 23.5$\pm$0.5 & 0.32$\pm$0.03 & 283$\pm$45 & 2.15 & 0.055  & 9.2$\pm$1.4  & 14  & 0.60  & -1.68$\pm$0.01 \\
Sextans	  	& 89	& -9.3$\pm$0.5	& 27.1$\pm$0.5 & 0.35$\pm$0.05 & 695$\pm$44 & 0.5  & 0.002  & 7.9$\pm$1.2  & 25  & 0.065 & -1.93$\pm$0.01 \\
\enddata
\tablenotetext{a}{Values taken from \citet{mcconnachie2012}}
\tablenotetext{b}{Values taken from \citet{mateo1998}}
\end{deluxetable*}

\citet{marks2011} used simulations to predict that the binary fraction should be larger for lower star formation rates. We use the time to form 50\% or 95\% of the stellar mass as a proxy for star formation rate and find the opposite --- binary fraction is higher for fast star formation rates. This discrepancy could be caused by possible invalid assumptions in the models by \citet[][i.e., that all stars start as members of binary systems]{marks2011}, or small number statistics and large error bars on our findings.

It has been suggested that a more densely populated star forming region should have a larger binary fraction \citep{kounkel2016}. We do not see this reflected in the mass density or luminosity density of the dwarfs, but this is likely because the properties we have access to do not translate to the density of progenitor star-forming regions. Overall, the quality and quantity of the data are not sufficient to discern any meaningful trends with binary fraction.

\section{Conclusions}\label{sec_conclusions}

Velocity data for Draco and Ursa Minor have been accumulating since the early 1980s \citepalias{olszewski1995,armandroff1995,kleyna2002,kleyna2003,wilkinson2004,kirby2010,walker2015}. We identified, collected, and combined the available data to produce the largest multi-epoch data set of radial velocities in both dwarfs. While many of these data sets have been used in previous studies to achieve a myriad of kinematic results, all of them required additional culling before we could use them for our purposes. The most involved process was for the \citetalias{walker2015} and Table \ref{tab:umi_table1} data, which entailed a maximum likelihood estimation of the velocity dispersion and systemic velocity that could be used for membership identification. 

This extensive velocity data made it possible for us to explore the binary populations in Draco and Ursa Minor. We generated MC simulations of the data and used a Bayesian technique that was developed in \citetalias{spencer2017b} and improved upon in this work to determine the binary fractions in Draco and Ursa Minor. By testing six different binary orbital parameter combinations for mass ratio and period, we conclude that the binary fraction for Draco is between $0.29^{+0.03}_{-0.03}$ and $0.69^{+0.07}_{-0.06}$, and the binary fraction for Ursa Minor is between $0.45^{+0.05}_{-0.05}$ and $0.96^{+0.03}_{-0.06}$. The most commonly used period and mass ratio distributions come from \citetalias{duquennoy1991}, which yielded binary fractions of $0.50^{+0.04}_{-0.06}$ and $0.78^{+0.08}_{-0.09}$ in Draco and Ursa Minor, respectively.

Changes to the shape of the period distribution had the largest effect on the posterior of the binary fraction, causing it to vary by as much as 30-50\%. The values we tested for the period distribution were inspired by observations of F--M type stars \citepalias{duquennoy1991,fischer1992} and simulations of stars in dwarf irregular galaxies \citepalias{marks2011}. The mass ratio distributions that we tested only produced binary fractions that varied by 4-11\%. Future work toward refining these distributions should focus on the period distribution, because it plays a larger role in determining the binary fraction. We found that the \citetalias{duquennoy1991} mass ratio distribution always did a better job of reproducing the data than the \citetalias{raghavan2010} distribution for a given period distribution. Period distributions peaking at $\mu_{\log P}=4.8$ or $\mu_{\log P}=5.8$ were always preferred over a distribution peaking at shorter periods ($\mu_{\log P}=3.5$).

Finally, we explored whether binary fraction is constant among dSphs by expanding our sample of two dwarfs to include Leo II, Carina, Fornax, Sculptor, and Sextans. We calculated the binary fraction for the additional dwarfs in the same way as was done for Draco and Ursa Minor, using velocity data from \citetalias{spencer2017b} for Leo II and from \citet{walker2009a} for Carina, Fornax, Sculptor, and Sextans. The probability that the binary fraction is constant (i.e., exists within a range of $f$ with width of 0.2 for all dwarfs) is $<1\%$, regardless of the inclusion of Fornax or the combination of period and mass ratio distributions. If binary fraction was a constant value, then the period distributions for each dwarf would need to vary. While this cannot be ruled out, it is certain that the binary populations within these dwarfs are different. That is to say, at least one property---be it binary fraction, period distribution, or something else---is not constant over all dwarfs.

Because we found that binary fraction varied given a fixed period distribution, we considered how binary fraction may vary with a variety of dSph properties. The strongest trends we found were that binary fraction was larger for dSphs that formed 50\% or 95\% of their stars faster and for dwarfs with larger velocity dispersions. Incorporating additional data for Carina, Fornax, Sculptor, and Sextans from other sources would allow for a better determination of their binary fractions and should yield cleaner trends with binary fraction if such trends exist.

\acknowledgements
The Hectochelle observations reported here were obtained at the MMT Observatory, a joint facility of the University of Arizona and the Smithsonian Institution.

The authors would like to thank Jan Kleyna and Mark Wilkinson for sharing their velocity data from \citetalias{kleyna2002}, \citetalias{kleyna2003}, and \citetalias{wilkinson2004} with us. We would also like to thank Josh Simon for allowing us to use his Keck spectra to obtain velocities for the \citetalias{kirby2010} data set. We warmly thank Andrew Szentgyorgyi, Nelson Caldwell, and the rest of the Hectochelle builders and staff. We also thank Marc Lacasse and the MMT operators, robot operators, and technical staff. We thank the anonymous referee for helpful comments that improved this work.

M.E.S. is supported by the National Science Foundation Graduate Research Fellowship under grant number DGE1256260. M.M. acknowledges support from NSF grants AST-1312997 and AST-1815403. M.G.W. acknowledges support from National Science Foundation grants AST-1313045 and AST-1813881. E.O. acknowledges support from NSF grants AST-1313006 and AST-1815767.

\end{document}